\documentclass[10pt,twocolumn]{article}
\usepackage[top=1.7cm, bottom=1.9cm, left=1.8cm, right=1.8cm]{geometry}
\usepackage{times}  %
\usepackage[sort&compress,numbers]{natbib}  %
\usepackage[hyphens]{url}
\usepackage{indentfirst}

\usepackage[hyphens]{url}  %
\usepackage{titling}
\usepackage[pdftex]{graphicx}
\graphicspath{{images/}}
\usepackage{amsmath,amsthm,amssymb,graphicx}
\usepackage{algorithm,algorithmicx,algpseudocode,verbatim,color}
\makeatletter
\algnewcommand{\LineComment}[1]{\Statex \hskip\ALG@thistlm \(\triangleright\) #1}
\makeatother
\makeatletter
\algnewcommand{\BlankLine}[1]{\Statex \hskip\ALG@thistlm #1}
\makeatother
\usepackage{array}
\usepackage{subcaption}
\usepackage[hyphens]{url}
\usepackage[all]{xy}
\usepackage[labelfont=bf]{caption}
\usepackage{multirow}
\usepackage{makecell}
\usepackage{pifont}%
\newcommand{\cmark}{\ding{51}}%
\newcommand{\xmark}{\ding{55}}%
\usepackage{longtable}
\usepackage[T1]{fontenc}
\usepackage{paralist}
\usepackage{xspace}
\usepackage{booktabs}
\usepackage[bookmarks=true, bookmarksnumbered=true, colorlinks=true, linkcolor=linkcol, citecolor=citecol, urlcolor=urlcol, hypertexnames=true]{hyperref}

\usepackage{framed}

\renewenvironment{quote}{%
  \list{}{%
    \leftmargin0.1cm
    \rightmargin\leftmargin
  }
  \item\relax
}
{\endlist}

\usepackage[marginal,hang,stable]{footmisc}
\renewcommand{\footnoterule}{
	\kern -3pt
	\hrule width 1in
	\kern 2pt
}

\usepackage{kantlipsum}
\usepackage[table, usenames, dvipsnames]{xcolor}
\usepackage[breakable, theorems, skins]{tcolorbox}
\tcbset{enhanced}
\newcommand{\charterfont}{\fontfamily{put}\fontsize{9pt}{11pt}\selectfont}

\newtcolorbox{takeawaybox}[1]{fonttitle=\bfseries,title=#1,colframe=red!45!black,fontupper=\charterfont\color{red!45!black},boxsep=3pt,left=5pt, right=5pt, top=5pt, bottom=5pt}  
\newcommand*\justify{%
  \fontdimen2\font=0.4em%
  \fontdimen3\font=0.2em%
  \fontdimen4\font=0.1em%
  \fontdimen7\font=0.1em%
  \hyphenchar\font=`\-%
}

\usepackage{etoolbox}

\renewcommand{\texttt}[1]{%
  \begingroup
  \ttfamily
  \spaceskip=0.2em
  \xspaceskip=0.2em
  #1%
  \endgroup
}

\usepackage[raggedright,compact]{titlesec}
\titlespacing*{\section}{0pt}{*2}{3pt}
\titlespacing*{\subsection}{0pt}{*2}{3pt}
\titlespacing*{\subsubsection}{0pt}{*2}{2pt}

\definecolor{darkgreen}{RGB}{0, 100, 0}
\definecolor{linkcol}{rgb}{0.3,0,0}
\definecolor{citecol}{rgb}{0.3,0,0}
\definecolor{urlcol}{rgb}{0.3,0,0}
\definecolor{vlightgray}{gray}{0.925}
\definecolor{teal}{rgb}{0.0, 0.5, 0.5}
\definecolor{pos}{HTML}{40B0A6}
\definecolor{neg}{HTML}{E1BE6A}
\definecolor{semipos}{HTML}{F0E442}

\newcommand{\descr}[1]{\smallskip\noindent\textbf{#1}}
\newcommand{\empeps}{\varepsilon_{\mathrm{emp}}}
\newcommand{\empdel}{\delta_{\mathrm{emp}}}
\newcommand{\puredp}{$\varepsilon$-DP\xspace}
\newcommand{\approxdp}{($\varepsilon, \delta$)-DP\xspace}
\newcommand{\rdp}{$(\alpha, \gamma)$-RDP\xspace}
\newcommand{\fdp}{$f$-DP\xspace}
\newcommand{\gdp}{$\mu$-GDP\xspace}

\newcommand{\FNR}{\text{FNR}}

\newif\ifcomment
\commenttrue
\ifcomment
	\newcommand{\edc}[1]{\textbf{\em\color{red}EDC: #1}}
	\newcommand{\sundar}[1]{\textbf{\em\color{blue}MS: #1}}
    \newcommand{\jh}[1]{\textbf{\em\color{teal}JH: #1}}
    \newcommand{\bb}[1]{\textbf{\em\color{magenta}BB: #1}}
    \newcommand{\gk}[1]{\textbf{\em\color{purple}GK: #1}}
\else
	\newcommand\edc[1]{}
	\newcommand\sundar[1]{}
    \newcommand\jh[1]{}
    \newcommand\bb[1]{}
    \newcommand\gk[1]{}
\fi

\newtheorem{theorem}{Theorem}[section]

\newtheorem{definition}{Definition}[section]

\usepackage{mathtools}
\newlength{\myeqskip}  
\AtBeginDocument{%
    \setlength\abovedisplayskip{\myeqskip}%
    \setlength\belowdisplayskip{\myeqskip}%
    \setlength\abovedisplayshortskip{\myeqskip-\baselineskip}%
    \setlength\belowdisplayshortskip{\myeqskip}}

\usepackage{enumitem}
\setlist[itemize]{leftmargin=1.5em, itemsep=0.75em, topsep=0pt, partopsep=2pt, parsep=0pt}

\usepackage{fourier}

\makeatletter
\renewenvironment{thebibliography}[1]
  {\section*{\refname}
   \small
   \begin{list}{\@biblabel{\@arabic\c@enumiv}}%
     {\setlength{\itemsep}{4pt}%
      \usecounter{enumiv}}}%
  {\end{list}}
\makeatother

    \newcommand{\edit}[1]{#1}
\newcommand{\descrit}[1]{\smallskip\noindent\textit{#1}}

\usepackage{microtype}

\title{\bf The Hitchhiker's Guide to Efficient, End-to-End,\\and Tight DP Auditing\thanks{Published at the 4th IEEE Secure and Trustworthy Machine Learning Conference (IEEE SaTML 2026) -- please cite the SaTML version.}}

\begin{document}

\sloppy

\author{Meenatchi Sundaram Muthu Selva Annamalai$^1$, Borja Balle$^2$, Jamie Hayes$^2$,\\Georgios Kaissis$^2$, and Emiliano De Cristofaro$^3$\\[1ex]
$^1$University College London, $^2$Google DeepMind, $^3$UC Riverside}
\date{}

\maketitle

 \begin{abstract}
In this paper, we systematize research on auditing Differential Privacy (DP) techniques, aiming to identify key insights and open challenges.
First, we introduce a comprehensive framework for reviewing work in the field and establish three cross-contextual desiderata that DP audits should target---namely, efficiency, end-to-end-ness, and tightness.
Then, we systematize the modes of operation of state-of-the-art DP auditing techniques, including threat models, attacks, and evaluation functions.
This allows us to highlight key details overlooked by prior work, analyze the limiting factors to achieving the three desiderata, and identify open research problems. 
Overall, our work provides a reusable and systematic methodology geared to assess progress in the field and identify friction points and future directions for our community to focus on.

 \end{abstract}

\section{Introduction}
\label{sec:intro}
Aiming to mitigate the privacy risks stemming from sharing or releasing sensitive datasets~\cite{sweeney2000simple,narayanan2008robust}, Differential Privacy (DP)~\cite{dwork2006calibrating} provides a robust mathematical framework offering formal privacy guarantees.
More precisely, DP bounds the impact that any single user can have on the output of a data release.
In recent years, DP techniques have been deployed in both the public and private sectors.
High-profile examples include the 2020 US Census release~\cite{abowd2022tda}, Apple's local DP system for learning user preferences~\cite{apple2017learning}, Google's next-word prediction models~\cite{mcmahan2022federated}, and Microsoft's synthetic data release with the UN's International Organization for Migration~\cite{microsoftiom}.%
\footnote{Refer to \url{https://desfontain.es/blog/real-world-differential-privacy.html} for a comprehensive list of real-world DP deployments.}

\begin{figure*}[t]
  \centering
  \includegraphics[width=0.72\linewidth]{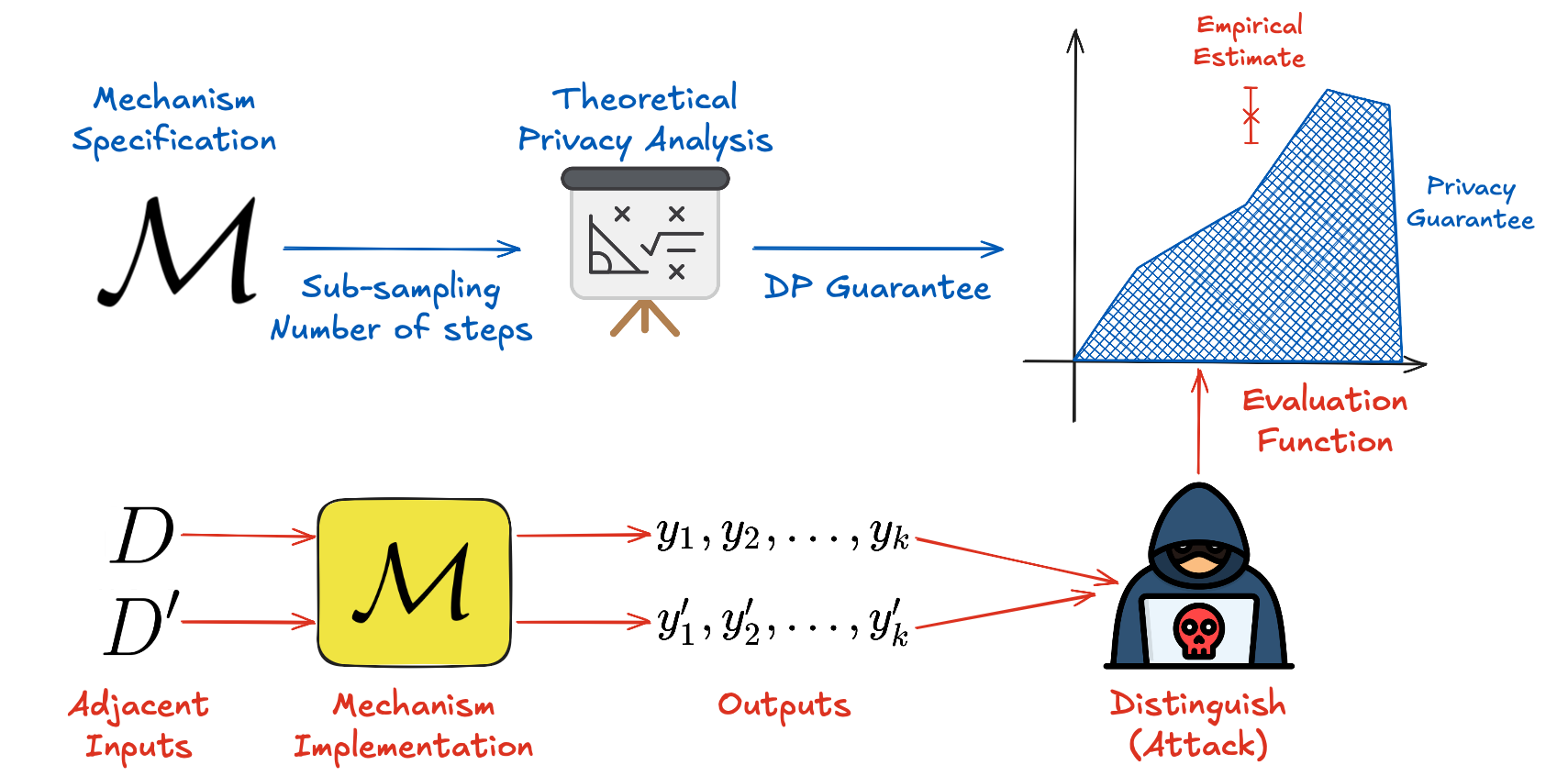}
  \caption{High-Level Overview of DP Auditing.}
  \label{fig:overview_audit}
  \vspace{-0.2cm}
\end{figure*}

\descr{DP Auditing.} In practice, implementing differentially private algorithms correctly can be quite challenging.
In the presence of incorrect implementations and/or bugs, the guarantees provided by DP can be severely degraded~\cite{nasr2023tight,annamalai2024shuffle} or completely compromised~\cite{tramer2022debugging,prngreuse}, resulting in the unintended leakage of sensitive user data. 
This has prompted extensive research on verifying whether DP's theoretical guarantees are met in practice~\cite{ding2018detecting,bichsel2021dp,jagielski2020auditing,nasr2023tight,lokna2023group,annamalai2024what,steinke2023privacy,xiang2025privacy}, a process known as {\em DP auditing}~\cite{kashyap2013testing}. 
As outlined in Figure~\ref{fig:overview_audit}, this involves conducting experiments on a mechanism's implementation, analyzing its outputs, and comparing against the theoretical guarantees of the mechanism's \emph{specification}.

DP auditing is a powerful and versatile technique to investigate the {\em tightness} of theoretical guarantees~\cite{jagielski2020auditing,nasr2021adversary,nasr2023tight,annamalai2024nearly,mahloujifar2024auditing}, develop theoretical intuitions for improved privacy analyses~\cite{nasr2025the,cebere2025tighter}, and refute conjectures~\cite{annamalai2024s}.
Moreover, it can be used to bound the adversarial success with respect to privacy attacks like membership inference~\cite{jagielski2020auditing}, attribute inference~\cite{malek2021antipodes}, or reconstruction~\cite{hayes2023bounding,mahloujifar2024auditing}.
Overall, DP auditing has important connections to many areas of privacy and security research, including DP verification, privacy accounting, etc.

\descr{Motivation.} While research on DP auditing dates back to at least 2013~\cite{kashyap2013testing}, extending and improving DP auditing techniques or proposing new ones still is a very active area of research.
Early work~\cite{ding2018detecting,bichsel2021dp} %
required running the implementations millions of times, making their computational cost prohibitively high for auditing complex private machine learning pipelines. 
Recent efforts have focused on {\em efficiency}, aiming to reduce the number of implementation runs down to a single one~\cite{steinke2023privacy,mahloujifar2024auditing,xiang2025privacy}.
However, for complex DP mechanisms, these techniques are largely ineffective in identifying bugs and violations as the audits fail to produce estimates of the empirical privacy leakage that are {\em ``close''} to the theoretical DP bounds---i.e., the audits are not {\em ``tight.''}

While some results, e.g.,~\cite{nasr2023tight}, provide audits that are both efficient and tight, they require aggressive modifications to the implementations, including adversarial assumptions that might not always be realistic in practice.
Thus, these techniques may fail to detect a large class of bugs or extend to other mechanisms; we refer to this characteristic as not being {\em ``end-to-end.''}
This prompts the need for a unified approach to building and evaluating DP auditing techniques that can simultaneously be efficient, end-to-end, and tight, arguably achieving the ``gold standard'' of DP auditing.

\descr{Roadmap.} In this paper, we undertake a systematic examination of the recent (and growing) literature on DP auditing across several key dimensions. 
Our objective is to distill new insights, highlight open research challenges, and delineate promising directions to improve auditing methodologies.
We formalize three desiderata for DP auditing: efficiency, end-to-end-ness, and tightness, and present a comprehensive framework based on them to assess progress within the field. 
Finally, we provide a structured systematization of the operational modes of existing DP auditing techniques, encompassing threat models, attack strategies, and evaluation functions.

Our analysis highlights important aspects that have been overlooked by prior work, as well as the key obstacles to achieving the three desiderata.
For instance, a major limitation to tighter audits is the absence of tight theoretical privacy analyses for sub-sampling schemes, such as shuffling.
We also show that, although recent audits have largely concentrated on the restrictive ``black-box'' threat model (where adversaries can only insert a single input sample and observe only the algorithm's final output), far less attention has been devoted to slightly stronger threat models, in which audits also fail to be tight.
Finally, while efficient and tight audits have recently been presented for simple mechanisms, extending these guarantees to more realistic and complex mechanisms like DP-SGD~\cite{abadi2016deep} remains a distant goal even under stronger threat models.

\descr{Summary of Contributions.} In short, in this work:
\begin{enumerate}[itemsep=2pt]
    \item We introduce a comprehensive framework to systematize state-of-the-art research on DP auditing with respect to its key aspects, namely the algorithmic foundations and the modes of operation involved.
 	\item We analyze the progress made by work in this field vis-\`a-vis three main desired properties, i.e., efficiency, end-to-end-ness, and tightness, which are important considerations for data and privacy practitioners when choosing DP auditing techniques to deploy.
	\item We discuss several insights and directions that future work could focus on to improve auditing techniques.
\end{enumerate}

\section{Background} %
\label{sec:background}

\subsection{Differential Privacy}
Differential Privacy (DP) provides a rigorous framework to control the privacy risk to individuals in a dataset. 
It ensures that the output of a computation, e.g., an answer to a statistical query or inference from a machine learning model, is not significantly affected by the presence/absence of any one individual's data---commensurate to a parameter $\epsilon$.
Over the years, a few variants of the DP definition have been proposed, as reviewed next.

\begin{definition}[\bf Pure DP~\cite{dwork2006differential}]
    \label{def:puredp}
    A randomized mechanism $\mathcal{M} : \mathcal{D} \rightarrow \mathcal{R}$ satisfies \puredp if, for any two adjacent datasets $D, D' \in \mathcal{D}$ and $S \subseteq \mathcal{R}$, it holds:
    \begin{equation*}
        \Pr[\mathcal{M}(D) \in S]  \leq \mathrm{e}^\varepsilon \Pr[\mathcal{M}(D') \in S]
    \end{equation*}
\end{definition}
\noindent Pure DP is defined by a single $\varepsilon$ parameter describing the upper bound on the probability an adversary can distinguish between outputs on two adjacent inputs (cf.~Section~\ref{sec:adj_defns}).

\descr{Approximate DP.}
\approxdp is relaxation of Pure DP that introduces a ``failure probability'' $\delta$, i.e., the probability that the \puredp guarantee does not hold.

\begin{definition}[Approximate DP~\cite{dwork2006our}]
    \label{def:approxdp}
    A randomized mechanism $\mathcal{M} : \mathcal{D} \rightarrow \mathcal{R}$ satisfies \approxdp if, for any two adjacent datasets $D, D' \in \mathcal{D}$ and $S \subseteq \mathcal{R}$, it holds:
    \begin{equation*}
        \Pr[\mathcal{M}(D) \in S]  \leq e^\varepsilon \Pr[\mathcal{M}(D') \in S] + \delta
    \end{equation*}
\end{definition}

\descr{R{\'e}nyi DP (RDP).}
\rdp captures a collection of \approxdp guarantees in one definition to more tightly analyze privacy guarantees of DP mechanisms.

\begin{definition}[R{\'e}nyi divergence~\cite{renyi1961measures}]
    \label{def:rdiv}
    For any two probability distributions, $P$ and $Q$, the R{\'e}nyi divergence of order $\alpha > 1$ is:
    \begin{align*}
        D_\alpha(P, Q) & \triangleq \frac{1}{\alpha - 1} \log \mathbb{E}_{x \sim Q} \left(\frac{P(x)}{Q(x)}\right)^\alpha
    \end{align*}
    \mbox{and} %
    $D_1(P, Q) = \lim_{\alpha \rightarrow 1}D_\alpha(P, Q) = \mathbb{E}_{x \sim P}\log \left(\frac{P(x)}{Q(x)}\right)$.
\end{definition}

\begin{definition}[R{\'e}nyi DP~\cite{mironov2017renyi}]
    \label{def:rdp}
    A randomized mechanism $\mathcal{M} : \mathcal{D} \rightarrow \mathcal{R}$ satisfies \rdp if, for any two adjacent datasets $D, D' \in \mathcal{D}$, it holds:
    \begin{equation*}
        D_\alpha(\mathcal{M}(D), \mathcal{M}(D')) \leq \gamma
    \end{equation*}
    where $D_\alpha$ is the R{\'e}nyi divergence of order $\alpha \geq 1$.
\end{definition}

\descr{Functional DP (\fdp).}
\fdp uses the hypothesis testing interpretation of DP~\cite{kairouz2015composition} (reviewed in Section~\ref{sec:hypo_testing_defn}) to bound the type II error achievable by an adversary aiming to distinguish between the outputs of a DP mechanism on adjacent inputs at a significance level of $\alpha$.
A specific instantiation of \fdp is Gaussian Differential Privacy (aka \gdp), for $f = T(\mathcal{N}(0, 1), \mathcal{N}(\mu, 1))$.

\begin{definition}[Trade-off function~\cite{dong2019gaussian}]
    For any two probability distributions on the same space, $P$ and $Q$, the trade-off function $T(P, Q): [0, 1] \rightarrow [0, 1]$ is defined as:
    \begin{equation*}
        T(P, Q)(\alpha) \triangleq \inf_{\phi} \{\beta_\phi: \alpha_\phi \leq \alpha\}
    \end{equation*}
    with the infimum taken over all rejection rules $\phi$ for which $\alpha_\phi$ and $\beta_\phi$ are the type I and type II errors, respectively.
\end{definition}

\begin{definition}[Functional DP~\cite{dong2019gaussian}]
    \label{def:fdp}
    A randomized mechanism $\mathcal{M} : \mathcal{D} \rightarrow \mathcal{R}$ satisfies \fdp if, for any two adjacent datasets $D, D' \in \mathcal{D}$, and $\alpha \in [0, 1]$ it holds:
    \begin{equation*}
        T(\mathcal{M}(D), \mathcal{M}(D'))(\alpha) \geq f(\alpha).
    \end{equation*}
\end{definition}

\subsection{Adjacency}
\label{sec:adj_defns}
In the above DP definitions, the notion of adjacent (aka neighboring) datasets is deliberately left generic as it varies depending on the setting.
We now review three common notions of adjacency: %

\begin{itemize}[itemsep=2pt]
\item {\bf\em Add/Remove:} corresponds to inserting or deleting a record from the dataset. 
It is also referred to as ``unbounded DP'' since the dataset size is not the same, i.e., $|D| = |D'| \pm 1$.

\item {\bf\em Edit:} replacing one record with another. 
It is also referred to as ``bounded DP'', i.e., $|D| = |D'|$.
Guarantees under edit adjacency are typically twice as strong as add/remove as one edit corresponds to one remove and one add.

\item {\bf\em Zero-out~\cite{kairouz2021practical}:} introduced to bridge guarantees provided under the add/remove and edit adjacencies and simplify theoretical privacy analyses.
Two datasets are zero-out adjacent if exactly one record in one dataset is replaced with a special \emph{zero-out record}, $\bot$ in the other.
\end{itemize}

\subsection{Sub-Sampling}
A common technique used to amplify a mechanism's DP guarantees and/or make DP compatible with modern machine-learning techniques is known as \emph{sub-sampling}.
More precisely, a mechanism is run iteratively on small, randomly chosen subsets of the input dataset, resulting in substantially improved privacy guarantees compared to running the mechanism on the full dataset.
This is referred to as ``privacy amplification by sub-sampling''~\cite{chaudhuri2006random} and depends on the specific sub-sampling scheme used.
We now review four common ones: %

\begin{itemize}[itemsep=2pt]
\item {\bf\em Poisson:} at each iteration, each record is chosen with some sampling probability $q$.

\item {\bf\em Sampling w/ (or w/o) replacement:} a batch of $B$ records is chosen, at each iteration, by sampling uniformly at random w/ (or w/o) replacement.
For brevity, we denote these schemes as Sampling WR and WOR, respectively.

\item {\bf\em Shuffling:} the dataset is randomly permuted (shuffled) and then partitioned into $b$ batches of $B$ records.
The batches are iterated over in sequence until the final one is processed, and the dataset is randomly permuted again.

\item {\bf\em Balls-in-bins~\cite{choquette-choo2025nearexact,chua2025ballsandbins}:} each record is uniformly assigned to one of $b$ batches, then, batches are processed in sequence for each iteration.
After the final batch is processed, the first one is processed again in a round-robin fashion.
\end{itemize}

\subsection{Hypothesis Testing Interpretation of DP}
\label{sec:hypo_testing_defn}

The notion of hypothesis testing interpretation of DP, introduced in~\cite{kairouz2015composition} for \approxdp, allows to evaluate a mechanism's privacy guarantees based on the difficulty an adversary faces in distinguishing between two adjacent datasets.
Given a random output from a mechanism $y \leftarrow \mathcal{M}(D^*)$ (where $D^* \in \{D, D'\}$ for adjacent $D, D'$), DP can be interpreted as inducing the following hypothesis test:
\begin{align*}
    H_0: y \sim \mathcal{M}(D) \quad H_1: y \sim \mathcal{M}(D')
\end{align*}

\noindent{\bf Privacy Region.}
For a choice of rejection region $S$, i.e., $H_0$ is rejected if $y \in S$, the type I \edit{(False Alarm)} and type II \edit{(Missed Detection)} errors are defined as $P_{\mathrm{FA}}(D, D', \mathcal{M}, S) = \Pr[\mathcal{M}(D) \in S]$ and $P_{\mathrm{MD}}(D, D', \mathcal{M}, S) = \Pr[\mathcal{M}(D') \notin S]$, respectively.
Thus, \approxdp bounds the type I and type II errors as per the following theorem:

\begin{theorem}[\cite{kairouz2015composition}]
    A randomized mechanism $\mathcal{M}: \mathcal{D} \rightarrow \mathcal{R}$ satisfies \approxdp if and only if the following conditions are satisfied for all adjacent datasets $D,D'$ and all rejection regions $S \subseteq \mathcal{R}$:
    \begin{align*}
        & P_{\mathrm{FA}}(D, D', \mathcal{M}, S) + \mathrm{e}^\varepsilon P_{\mathrm{MD}}(D, D', \mathcal{M}, S) \geq 1 - \delta\; \land \\
        & \mathrm{e}^\varepsilon P_{\mathrm{FA}}(D, D', \mathcal{M}, S) + P_{\mathrm{MD}}(D, D', \mathcal{M}, S) \geq 1 - \delta
    \end{align*}
\end{theorem}

This interpretation can also be operationalized, i.e., an analyst can pick a pair of adjacent datasets $D, D'$ and plot the type I and type II errors obtained for different rejection regions $S$.
This will form a ``privacy region'' defined by \approxdp, which %
is often used for DP auditing. 

\subsection{Related Work}
\label{sec:related}
To the best of our knowledge, our work is the first to systematically review and analyze the body of work on DP auditing.
In the following, we briefly review other systematization-of-knowledge and survey papers on related topics.
Namatevs et al.~\cite{namatevs2025privacy} recently present a survey of DP auditing research; however, they do not analyze relevant work within a systematic framework.
By contrast, we do so to comprehensively analyze the state of the art, measure progress, identify gaps and open research problems primarily along three key characteristics: whether audits are efficient, end-to-end, and tight.

Systematization of knowledge efforts in related areas includes the work by Papernot et al.~\cite{papernot2018sok}, who present attacks and defenses on (non-private) machine learning systems using a comprehensive adversarial framework.
Salem et al.~\cite{salem2023sok} focus on privacy risks and provide a unifying framework relying on privacy games.
Finally, Desfontaines and Pejó~\cite{desfontaines2020differential} systematize the various variants and extensions of DP.
While DP auditing relies on and is informed by the notions of privacy attacks, games, and DP variants, our work focuses on the specific process of deriving empirical bounds, which has thus far not been explored in prior work.

\section{DP Auditing}
\label{sec:dp_audit_defn}
DP auditing denotes a set of empirical techniques involving an adversary instantiated against a DP mechanism to distinguish between outputs on adjacent datasets.
We now discuss its two main motivating use cases and the desired properties we believe DP auditing algorithms should satisfy.

\subsection{Goals of DP Auditing}
\label{sec:goals_of_auditing}

Arguably, the two main goals of DP auditing include: 1) \emph{verifying} that the theoretical guarantees provided by DP are met in practice and 2) \emph{estimating} the actual privacy leakage observed from the mechanism empirically.

\descr{Verifying Privacy Guarantees.}
While increased adoption of DP in real-world applications~\cite{abowd2022tda,apple2017learning,microsoftiom,mcmahan2022federated,tumult2025tumult,opendp2025opendp} is a promising development for privacy-enhancing technologies, it can often be difficult to implement DP correctly.
Bugs leading to unintended privacy leakage, in some cases completely breaking DP guarantees, have been found in many implementations~\cite{lyu2017understanding,zhang2016privtree,prngreuse,lokna2023group,annamalai2024what,annamalai2024shuffle,ganev2025elusive}.
For instance, Lyu et al.~\cite{lyu2017understanding} and Zhang et al.~\cite{zhang2016privtree} focus on the differentially private Sparse Vector Technique (SVT)~\cite{dwork2009complexity} and show, through manual analysis, that most of its implementations~\cite{roth2011lecture,dwork2014algorithmic,lee2014top,stoddard2014differentially,chen2015differentially,lyu2017understanding} are not differentially private at all.

Bugs in the implementation of machine-learning algorithms like Differentially Private Stochastic Gradient Descent (DP-SGD)~\cite{abadi2016deep,prngreuse,annamalai2024shuffle} and DP synthetic data~\cite{annamalai2024what,ganev2025elusive} have also been found through both manual and automated techniques.
However, manual analysis can be error-prone and may miss subtle implementation details, thus prompting the need to scale up automated approaches~\cite{annamalai2024what}.

\descrit{Limitations of Static Analysis.}
Early work aiming to automatically verify DP guarantees in algorithms' implementations has used static code analysis tools, e.g., type checkers and specialized programming languages.
For example, Reed and Pierce~\cite{reed2010distance} present a type system that enables programmers to write privacy-safe programs by design.
Overall, a large body of work~\cite{barthe2012probabilistic,gaboardi2013linear,barthe2013verified,barthe2015higher,zhang2017lightdp,albarghouthi2017synthesizing,wang2019proving,wang2020checkdp} has focused on making type systems more expressive and efficient.

However, this approach faces some fundamental limitations.
First, type systems implicitly trust that some ``basic'' DP algorithms are implemented correctly by the underlying programming language (e.g., Gaussian mechanism, Laplace mechanism).
As pointed out in~\cite{lokna2023group}, this may not necessarily be the case if, e.g., the noise addition mechanisms suffer from floating-point vulnerabilities~\cite{mironov2012significance}.
Second, these type systems require mechanisms to be implemented in a very specific language, thus, they are substantially limited to simple algorithms like SVT or Report Noisy Max~\cite{dwork2014algorithmic}. %
Specifically, they struggle to capture the complexities involved in algorithms like RAPPOR~\cite{erlingsson2014rappor} or DP-SGD~\cite{abadi2016deep},

\descrit{DP Auditing's Role.}
Rather than statically checking %
program specifications, %
DP auditing allows an auditor to run experiments and {\em statistically} refute the theoretical privacy guarantees provided by DP.
In other words, if the auditor fails, programmers and clients can be assured that the implementation satisfies the DP guarantees at some (possibly pre-defined) level of confidence---e.g., for more critical applications, the auditor can perform the test at 99.999\% confidence~\cite{tramer2022debugging} or 95\% for less critical applications~\cite{jagielski2020auditing}.
Overall, DP audits can offer more flexibility than static methods, accommodating more complex algorithms such as private machine learning pipelines~\cite{jagielski2020auditing}.
They are also easier to implement as they do not involve specific programming languages or typing rules.
\descr{Empirical Privacy Estimation.}
DP auditing can also provide precise estimates of the privacy leakage in different settings, e.g., vis-\`a-vis different real-world adversaries and/or attacks.

\descrit{From Worst-Case to Real-World Adversaries.}
DP provides formal privacy guarantees against powerful theoretical adversaries that are conceivably not always realistic in practice.
In some cases, these strong adversarial assumptions may be needed to provide robust protections that hold in worst-case settings.
In other cases, this may instead be a direct consequence of limited known ways of proving privacy guarantees~\cite{nasr2021adversary}.
For instance, the privacy analysis of DP-SGD~\cite{abadi2016deep} assumes an adversary that has access not only to the final trained model but also to all intermediate models at each training step.
This assumption is not entirely realistic across the board as intermediate parameters are not usually released, %
but needed since theoretical privacy analysis capabilities in this threat model are limited~\cite{feldman2018privacy,chourasia2021differential,ye2022differentially}.
Therefore, DP auditing can be used to empirically estimate privacy leakage in real-world settings~\cite{jagielski2020auditing,annamalai2024s,annamalai2024nearly,nasr2025the,cebere2025tighter}.

In this context, an empirical estimate $\empeps$ can be derived experimentally and then compared against the theoretical bound $\varepsilon$.
This allows to investigate the {\em tightness} of theoretical guarantees~\cite{jagielski2020auditing,nasr2021adversary,nasr2023tight,annamalai2024nearly,mahloujifar2024auditing}, develop theoretical intuitions for improved privacy analyses~\cite{nasr2025the,cebere2025tighter}, and refute widely believed conjectures~\cite{annamalai2024s}.
 
\descrit{Testing Resistance to Different Privacy Attacks.} 
Another use-case of DP auditing stems from its modularity.
DP auditing can be instantiated, e.g., on different adjacency definitions or DP variants or using different attacks.
While the DP definition directly maps to the notion of membership inference (MI), which we review in Section~\ref{sec:attack_defns}, DP guarantees can also be extended to and empirically verified against other attacks such as attribute inference (AI) and reconstruction attacks (Recon)~\cite{yeom2020overfitting,salem2023sok,hayes2023bounding}. 
Privacy analyses under more general privacy attacks may not always be optimal, thus, DP auditing can be used to investigate the tightness of the protections provided by DP against them~\cite{hayes2023bounding,mahloujifar2024auditing}.

\subsection{DP Auditing Algorithms}
We now present a formal definition of a DP auditing algorithm, which we use in our systematization to identify relevant papers in this area. %

To ease our definition, we first define an ordering over the space of possible DP guarantees $\mathcal{K}$:
\begin{definition}[Ordering of DP Guarantees]\label{def:dp_order}
    For $\kappa, \kappa' \in \mathcal{K}$, $\kappa \succeq \kappa'$ if $\mathcal{M}$ satisfies $\kappa'$-DP $\rightarrow$ $\mathcal{M}$ satisfies $\kappa$-DP.
\end{definition}
\noindent In Pure DP, $\mathcal{K} = \mathbb{R}^+ \cup \{0\}$, $\kappa = \varepsilon$, and the ordering follows the natural ordering of real numbers.

\begin{definition}[DP Auditing Algorithm]\label{def:dp_audit_alg}
    A DP auditing algorithm $\mathcal{A}_{\mathrm{DP}}$ takes in input a $\kappa$-DP mechanism $\mathcal{M}$, the claimed DP guarantees $\tilde{\kappa} \in \mathcal{K}$, and some additional hyper-parameters $\Theta$ and outputs $z \leftarrow \mathcal{A}_{\mathrm{DP}}(\mathcal{M}, \tilde{\kappa}; \Theta)$.\\[0.5ex]
    If $\mathcal{A}_{\mathrm{DP}}$ is used for verification, $z \in \{0, 1\}$ and $\mathcal{A}_{\mathrm{DP}}(\mathcal{M}, \tilde{\kappa}; \Theta) = 1$ if $\tilde{\kappa} \succeq \kappa$. If used for empirical estimation, $z \in \mathcal{K}$ and $\kappa \succeq \mathcal{A}_{\mathrm{DP}}(\mathcal{M}, \tilde{\kappa}; \Theta)$.
\end{definition}

The above definition also encodes a notion of ``correctness'' for DP auditing algorithms, which informally states that the output of the algorithm agrees with the actual guarantees satisfied by the mechanism (i.e., if $\mathcal{M}$ satisfies $\kappa$-DP, the algorithm confirms this to be true as well).
However, this definition admits vacuous algorithms (e.g., an algorithm that always outputs `1'); thus, in Section~\ref{sec:desiderata}, we define additional desirable properties of DP auditing algorithms.

\descr{Auditing Game.} A DP mechanism is typically audited vis-\`a-vis a \emph{privacy game} played between an \textit{Adversary} and a \textit{Challenger}.
Although this game can slightly vary depending on the auditing goal and threat model, we present the prototypical DP audit algorithm in Algorithm~\ref{alg:audit_game}.

First, the Adversary chooses an initial dataset in the data domain $D^-$
and distributions from which special data points used to audit the mechanism, i.e., \emph{canaries}, are sampled from. 
They can also provide ``auxiliary'' information to the mechanism if necessary.

Next, the Challenger runs the mechanism $n_{\mathrm{R}}$ times, each time sampling $n_{\mathrm{C}}$ fresh canaries from the canary distributions to form the input dataset.
The Challenger then sends the outputs back to the Adversary,
who post-processes the outputs of the mechanism, typically using a privacy attack $\mathcal{P}$.
This is because the outputs of the mechanism may be in an arbitrary domain $y \in \mathcal{Y}$, which can be difficult to design a decision function around (e.g., to decide if the mechanism satisfies the claimed guarantees).
Thus, the post-processing yields simple observations that can be easy to analyze.

Finally, the outcome of the audit can be computed from the observations using an evaluation function $\zeta$, which outputs either a bit $z \in \{0, 1\}$ for the verification of a theoretical DP guarantee or $z \in \mathcal{K}$ for empirical privacy estimation.
This is typically done by relying on the hypothesis testing interpretation of DP discussed above.
In practice, the outcome of the audit is typically computed with a desired level of confidence $\alpha$, especially when refuting/confirming the claimed DP guarantees of the mechanism.

\begin{algorithm}[t]
    \small
    \caption{Prototypical Auditing Algorithm}\label{alg:audit_game}
    \begin{algorithmic}[1]
    \Require Mechanism, $\mathcal{M}$. DP Guarantee, $\kappa \in \mathcal{K}$. Data domain, $\mathcal{X}$. Number of runs, $n_{\mathrm{R}}$. Number of canaries, $n_{\mathrm{C}}$. Confidence level, $\alpha$. Privacy attack, $\mathcal{P}$. Evaluation function, $\zeta$.
    \LineComment \textbf{Adversary}
    \State \textbf{Pick dataset:} $D^- \in \mathcal{X}^*$
    \State \textbf{Pick canary distributions:} $\mathcal{C}$
    \State \textbf{Define auxiliary information:} $\mathtt{aux}$

    \LineComment \textbf{Challenger}
    \State $Y = [\;]$
    \State $D_{\mathrm{C}} = [\;]$
    \For{$i \in [n_{\mathrm{R}}]$}
        \State $D_{\mathrm{C}}[i] \leftarrow \{x \sim \mathcal{C}[i] | i \in [n_{\mathrm{C}}]\}$
        \State $Y[i] \leftarrow \mathcal{M}(D^- \cup D_{\mathrm{C}}[i]; \mathtt{aux})$
    \EndFor

    \LineComment \textbf{Adversary}
    \State $O = [\;]$
    \For{$i \in [n_{\mathrm{R}}]$}
        \State $O[i] \leftarrow \mathcal{P}(Y[i]; \mathcal{M}, D^-, \mathcal{C})$
    \EndFor

    \State \Return $z \leftarrow \zeta(O; \mathcal{M}, D_{\mathrm{C}}, \mathcal{C}, \alpha, \kappa)$
    \color{black}
    \end{algorithmic}
\end{algorithm}

\subsection{Desiderata: The Gold Standard of DP Auditing}
\label{sec:desiderata}
Next, we define three key properties we believe audits should satisfy -- \emph{efficiency, end-to-end-ness, tightness}. %
We believe that achieving these three properties simultaneously would be optimal for the widespread and rigorous adoption of DP auditing, especially in production systems, although this remains an open research problem.
Even when they cannot all be achieved simultaneously, we believe that analyzing DP audits along these three dimensions is crucial in helping data and privacy practitioners choose the optimal auditing techniques for their specific contexts.

\descr{Efficiency.}
For DP audits to be deployable in practice, they have to be computationally efficient.
Early work~\cite{ding2018detecting,bichsel2018dp} required implementations to be executed millions of times -- while this is achievable for simple mechanisms (e.g., Laplace and Gaussian mechanisms), it can be prohibitively expensive for complex mechanisms like DP-SGD~\cite{abadi2016deep}.
Therefore, recent work~\cite{steinke2023privacy,mahloujifar2024auditing,galen2024oneshot,xiang2025privacy} has focused on developing auditing techniques that only require implementations to be executed a handful of times, possibly even just once.

\begin{definition}[Efficient Audit]
    A DP auditing algorithm $\mathcal{A}_{\mathrm{DP}}(\mathcal{M}, \kappa; \Theta)$ is efficient if it requires the mechanism $\mathcal{M}$ to be run only a few times\footnote{Although ``few'' can be context-dependent, in this work, we consider ``few'' to mean once or twice.}.
\end{definition}

\descr{End-to-End.} DP audits should also apply to {\em end-to-end} privacy guarantees of the implementations; ideally, these should be run by simply providing inputs to the mechanism and observing its final output.
This is often referred to as ``black-box'' auditing~\cite{liu2019minimax} and reflects threat models in practical settings as opposed to worst-case threat models assumed by the privacy analysis of certain mechanisms.
For instance, the privacy analysis of the DP-SGD~\cite{abadi2016deep} algorithm assumes that all intermediate models are released, even though in practice only the final model may be accessible to the adversary.
Since ``black box'' may have different meanings in different areas (e.g., query-access to machine learning models~\cite{shokri2017membership} vs.~access to fixed generated synthetic data records~\cite{houssiau2022tapas}, etc.), we use the notion of ``end-to-end'' auditing to avoid confusion.

End-to-end audits are also broadly applicable to many mechanisms as they do not require access to \emph{how} the mechanism is implemented internally.
For example, although different DP mechanisms can be used to train private ML models (e.g., DP-SGD~\cite{abadi2016deep}, PATE~\cite{papernot2017semisupervised}, DP-ZO~\cite{tang2025private}), relying on one single auditing technique that takes in input the training dataset and observe the trained model's outputs would make it significantly easier to integrate auditing into different training pipelines.
Once again, this would also more closely match the capabilities of realistic adversaries. %

\begin{definition}[End-to-End Audit]
    A DP auditing algorithm $\mathcal{A}_{\mathrm{DP}}(\mathcal{M}, \kappa; \Theta)$ is end-to-end if it does not modify the mechanism $\mathcal{M}$ and performs the audit simply by observing the mechanisms output(s) on specific input(s).
\end{definition}

\descr{Tightness.}
Informally, a \emph{tight} audit means that the empirical audit is consistent with the theoretical privacy analysis.
More precisely, the empirical estimates obtained through auditing should not be far from the theoretical guarantees (e.g., $\empeps \approx \varepsilon$ and $\empdel \approx \delta$) -- see Definition~\ref{def:tight}.

When verifying DP guarantees, tight DP audits ensure that implementations are verified only if their actual privacy leakage is less than or equal to the theoretical DP bounds.
Consequently, tight audits are required in this context to ensure accuracy and provide strong confidence in the process.
By contrast, a \emph{loose} audit that successfully verifies implementations in which the actual privacy leakage is, say, twice the theoretical upper bound would inherently not be very useful for identifying bugs. %

In theory, the definition below can be proved asymptotically or demonstrated empirically.
All papers we consider generally report empirical estimates of $\kappa$-DP, along with a confidence interval.
Therefore, we consider an audit ``tight'' if the theoretical guarantee lies within the confidence interval of the estimate with a ``sufficiently small'' confidence interval.

\begin{definition}[Tight Audit]\label{def:tight}
    A DP auditing algorithm $\mathcal{A}_{\mathrm{DP}}(\mathcal{M}, \kappa; \Theta)$ is tight if it it never underestimates the true privacy parameters, i.e., assuming $\mathcal{M}$ satisfies $\kappa$-DP, $\mathcal{A}_\mathrm{DP}(\mathcal{M}, \kappa; \Theta) = \kappa$ for estimation and $\forall \kappa' \succeq \kappa\; \mathcal{A}_\mathrm{DP}(\mathcal{M}, \kappa'; \Theta) = 0$ for verification.
\end{definition}


\newcommand{\myrule}{\specialrule{0.03em}{0.15em}{0.25em}}
\newcommand{\myspan}{\multirow{2}{*}}

\begin{table*}[t]
\small
		\setlength{\tabcolsep}{7.5pt}
    \begin{tabular}{@{}lcc@{}cc@{}cc@{}c@{}c@{}}
        \toprule
        & & \multicolumn{2}{c}{{\bf Foundations}} & \multicolumn{2}{c}{{\bf Operational Details}} & \multicolumn{3}{c}{{\bf Progress}} \\
        \cmidrule(l){3-4}        \cmidrule(l){5-6}         \cmidrule(l){7-9}
        {\bf Reference} & \makecell{{\bf Mechanisms}\\{\bf Audited}} & {\bf DP Guarantee~} & {\bf Sub-Sampling} & {\bf Attack} & \makecell{{\bf Eval.}\\{\bf Function}} & {\bf Efficient~} & {\bf End-to-End~} & {\bf Tight} \\
        \midrule
        Ding et al.~(2018)~\cite{ding2018detecting} & Simple & \puredp & -- & -- & Output-set & \xmark & \cmark & \cmark \\
        \myrule
        Bischel et al.~(2018)~\cite{bichsel2018dp} & Simple & \puredp & -- & -- & Output-set & \xmark & \cmark & \cmark \\
        \myrule
        Jagielski et al.~(2018)~\cite{jagielski2020auditing} & DP-SGD & \approxdp & Poisson & MI & FPR/FNR & \xmark & \cmark & \xmark \\
        \myrule
        \myspan{Nasr et al.~(2021)~\cite{nasr2021adversary}} & \myspan{DP-SGD} & \myspan{\approxdp} & \myspan{Poisson} & \myspan{MI} & \myspan{FPR/FNR} & \xmark & \cmark & \xmark \\
        & & & & & & \xmark & \xmark & \cmark \\
        \myrule
        Malek et al.~(2021)~\cite{malek2021antipodes} & ALIBI & Label DP & -- & AI & Accuracy & \cmark & \cmark & \xmark \\
        \myrule
        Houssiau et al.~(2022)~\cite{houssiau2022tapas} & Synthetic Data & \approxdp & -- & AI & FPR/FNR & \xmark & \cmark & \xmark \\
        \myrule
        Askin et al.~(2022)~\cite{askin2022statistical} & Simple & \puredp & -- & -- & Distance Est. & \xmark & \cmark & \cmark \\
        \myrule
        Lokna et al.~(2023)~\cite{lokna2023group} & Simple & \approxdp & -- & DPD~ & Output-set & \xmark & \cmark & \cmark \\
        \myrule
        \myspan{Nasr et al.~(2023)~\cite{nasr2023tight}} & \myspan{DP-SGD} & \myspan{\fdp} & Poisson & \myspan{MI} & \myspan{FPR/FNR} & \xmark & \cmark & \xmark \\
        & & & -- & & & \cmark & \xmark & \cmark \\
        \myrule
        Maddock et al.~(2023)~\cite{maddock2023canife} & DP-FedSGD & User-level DP & Poisson & MI & FPR/FNR & \cmark & \xmark & \xmark \\
        \myrule
        \myspan{Pillutla et al.~(2023)~\cite{pillutla2023unleashing}} & \myspan{DP-SGD} & \myspan{\approxdp} & \myspan{Poisson} & \myspan{MI} & \myspan{Custom} & \xmark & \cmark & \xmark \\
        & & & & & & \xmark & \xmark & \xmark \\
        \myrule
        \myspan{Steinke et al.~(2023)~\cite{steinke2023privacy}} & \myspan{DP-SGD} & \myspan{\approxdp} & \myspan{Poisson} & \myspan{MI} & \myspan{Accuracy} & \cmark & \cmark & \xmark \\
        & & & & & & \cmark & \xmark & \xmark \\
        \myrule
        Galen et al.~(2024)~\cite{galen2024oneshot} & DP-FedAvg & User-level DP & Shuffle & MI & Variance Est. & \cmark & \xmark & \xmark \\
        \myrule
        Chadha et al.~(2024)~\cite{chadha2024auditing} & PATE & \rdp & -- & MI & Custom & \xmark & \xmark & \xmark \\
        \myrule
        \myspan{Annamalai et al.~(2024)~\cite{annamalai2024what}} & \myspan{Synthetic Data} & \myspan{\approxdp} & \myspan{Poisson} & \myspan{MI} & \myspan{FPR/FNR} & \xmark & \cmark & \xmark \\
        & & & & & & \xmark & \xmark & \cmark \\
        \myrule
        Feng et al.~(2024)~\cite{feng2024privacy} & DP-SGD & \approxdp & Poisson & -- & Output-set & \xmark & \cmark & \cmark \\
        \myrule
        Annamalai et al.~(2024)~\cite{annamalai2024nearly} & DP-SGD & \gdp & -- & MI & FPR/FNR & \xmark & \cmark & \xmark \\
        \myrule
        \multirow{2}{*}{Mahloujifar et al.~(2024)~\cite{mahloujifar2024auditing}\hspace*{-0.5cm}}  & DP-SGD & \fdp & Poisson & \multirow{2}{*}{Recon~} & \multirow{2}{*}{Accuracy}  & \cmark & \xmark & \xmark \\
        & Gaussian & \gdp & -- &  &  & \cmark & \cmark & \cmark \\
        \myrule
        Annamalai et al.~(2024)~\cite{annamalai2024shuffle} & DP-SGD (Shuffle) & \approxdp & Shuffle & MI & FPR/FNR & \xmark & \xmark & \xmark \\
        \myrule
        Cebere et al.~(2025)~\cite{cebere2025tighter} & DP-SGD & \gdp & Custom & MI & FPR/FNR & \xmark & \cmark & \xmark \\
        \myrule
		\myspan{Xiang et al.~(2025)~\cite{xiang2025privacy}} & DP-SGD & \fdp & Poisson & \myspan{MI} & \myspan{Accuracy} & \cmark & \xmark & \xmark \\
		 & Gaussian & \gdp & -- & & & \cmark & \cmark & \cmark \\
        \bottomrule
    \end{tabular}
    \small
    \centering
    \caption{Summary of key prior work on DP auditing (see Appendix~\ref{sec:full_summary} for a full summary). NB: ``Simple'' mechanisms refer to Laplace, Report Noisy Max, Noisy Histogram, and Sparse Vector Technique mechanisms. Please see Sections~\ref{sec:background},~\ref{sec:desiderata}, and~\ref{sec:towards_sys} for definitions of DP guarantees and sub-sampling methods, progress metrics, and attacks and evaluation functions, respectively.}
    \label{tab:summary_intro}
    \vspace{-0.2cm}
\end{table*}

\section{The Current Body of Work on\\ DP Auditing}
\label{sec:body_of_work}
We now present our methodology to compile a comprehensive and representative snapshot of DP auditing research.
\descr{Querying Scopus.} 
In January 2025, we queried the Scopus research database~\cite{scopus} using several keywords (see Appendix~\ref{sec:exact_search} for the complete list) and obtained a total of 179 papers that mention DP auditing or similar terms in their title or abstract.
We use Scopus as it indexes the top-four Security and Privacy venues (IEEE S\&P, ACM CCS, USENIX, and NDSS), the major Machine Learning conferences (e.g., NeurIPS, ICLR, ICML), as well as prominent journals like the Transactions of Machine Learning Research (TMLR).

\descr{Manual Review.} We then manually reviewed the abstracts of the 179 articles and filtered relevant ones that either explicitly estimate the DP parameters of an implementation or verify whether claimed DP guarantees are met following Definition~\ref{def:dp_audit_alg}.
We also followed a snowballing approach (aka citation chaining) by examining the references cited in these papers to
identify recently accepted work on DP auditing not yet indexed by Scopus, as well as non-peer-reviewed papers that already informed published research at top conferences (e.g., articles published on arXiv).
This yielded an additional 24 papers.

\descr{Final List.} After manual review, we identified 45 relevant DP auditing papers, which form the basis for our systematization effort.
(Note that prior SoKs in adjacent fields typically systematize 25 to 30 papers~\cite{du2025sok,noppel2024sok,meeus2025sok}.)
To ease presentation, in Table~\ref{tab:summary_intro}, we report 21 key papers (out of the 45) that arguably represent the state of the art, while reporting earlier/seminal work in Appendix~\ref{sec:full_summary}. %
Specifically, Table~\ref{tab:summary_intro} lists papers that are the first to audit a particular DP guarantee (e.g.~\cite{ding2018detecting,nasr2023tight}), mechanism (e.g.~\cite{jagielski2020auditing,malek2021antipodes}), or achieve the current ``best'' auditing results (e.g.,~\cite{nasr2023tight,xiang2025privacy}) and are typically published at top-tier venues.
Along with the relevant selected papers, we also report key details of the auditing methods introduced in these papers, which we extract through our systematization process as discussed later in Section~\ref{sec:towards_sys}.

At first glance, Table~\ref{tab:summary_intro} reveals that state-of-the-art DP auditing spans a wide range of mechanisms, from ``simple'' ones (e.g., Laplace, Noisy Histogram, etc.) to more ``complex'' machine learning pipelines, vis-\`a-vis various DP guarantees and adjacency notions and employing different sub-sampling techniques. %
Consequently, to support this wide range of mechanisms and techniques, auditing algorithms use different privacy attacks and evaluation functions, which we analyze in Sections~\ref{sec:attack_defns} and~\ref{sec:eval_fns_defns}, respectively.
Through our systematization, we find that DP auditing techniques have met the three desiderata discussed in Section~\ref{sec:desiderata} only for simple mechanisms (e.g., Gaussian), whereas achieving them for complex mechanisms remains elusive.

Although our selection was done in January 2025, we are confident it is meaningfully representative of the state of the art in DP auditing as of the time of submission; i.e., even if one were to add or remove a small number of papers to our list, our main takeaways would not substantially change.

\descr{Out-of-scope Research.} As our systematization focuses on DP auditing, we do not include work estimating privacy leakage using privacy metrics other than DP guarantees, e.g., using attack success rate, TPR@low FPR, and/or adversarial advantage~\cite{pyrgelis2018knock,jayaraman2019evaluating,busa2024auditing,aerni2024evaluations,zarifzadeh2024lowcost,wen2023canary}.
We also excluded research focusing on using static methods, such as type checkers, to verify the DP guarantees of an implementation without executing the program.
While these methods are useful in various ways, they are arguably limited in the range of mechanisms they can audit, which makes them ineffective for verifying the DP guarantees of complex modern DP mechanisms, e.g., in the context of deep learning, as discussed in Section~\ref{sec:goals_of_auditing}.

\descr{DP-SGD.}
Although in recent years, much of the focus on DP auditing has been on DP-SGD~\cite{abadi2016deep}, our systematization takes a broader view of DP auditing as even for ``simple'' mechanisms, tight, end-to-end, and efficient have only been shown to be possible recently~\cite{mahloujifar2024auditing,xiang2025privacy}.
Furthermore, many of the novel audit methodologies that have been introduced for DP-SGD are modular in principle and can be applied to any mechanism in general~\cite{steinke2023privacy,mahloujifar2024auditing,xiang2025privacy}.
Therefore, in this work, we derive findings and recommendations that are broadly applicable to any DP mechanism in general.

\section{Systematizing DP Auditing Research Based on their Operational Modes}
\label{sec:towards_sys}
In this section, we systematize the state of the art on DP auditing based on the modes of operation specific to it. %
Specifically, we extract the threat models used in the literature and introduce a new taxonomy to further break down these threat models into specific adversarial capabilities.
Furthermore, we extract the types of privacy attacks and evaluation functions instantiated by auditing techniques.
Finally, we discuss different techniques  to calculate confidence levels for audits.

\subsection{Threat Modeling}
\label{sec:threat_model_defn}

When auditing DP implementations, especially those deployed in the wild, it is important to do so vis-\`a-vis meaningful threat models.
Naturally, these do not only affect the privacy guarantees of a mechanism but also impact whether the audit can be considered end-to-end as stronger adversarial models often require mechanisms to be modified before auditing.
In this section, we focus on auditing the DP-SGD algorithm~\cite{abadi2016deep} since the natural threat model for simple mechanisms like Laplace is already the {\em end-to-end} model. %

In the context of private machine learning, different threat models have been used to instantiate privacy attacks (e.g., black-box, white-box, etc.).
For more fine-grained analysis, we break them down into specific capabilities, i.e., model visibility and types of canary used by the attacks. This enables us to systematize research along these axes independently.

\descr{Model Visibility.}
When training ML models using DP-SGD, only the final trained model is released and visible to the adversary.\footnote{Except for settings like Federated Learning~\cite{mcmahan2017learning}, where algorithms like DP-FedAvg and DP-FedSGD inherently disclose intermediate model updates.}
However, the theoretical privacy analysis of DP-SGD assumes that the intermediate models across all iterations are released as we do not know how to prove DP guarantees otherwise.
Thus, early work assumes that the adversary has access to all intermediate models~\cite{nasr2021adversary,nasr2023tight}.
More recently, the focus has shifted to auditing DP-SGD when only the final model output~\cite{pillutla2023unleashing,steinke2023privacy,annamalai2024nearly,mahloujifar2024auditing} is available.

When only the final model is visible, the prototypical auditing algorithm outlined in Algorithm~\ref{alg:audit_game} should be modified so that the Challenger does not send all the intermediate outputs produced by the mechanism to the Adversary (line 8), but rather only the final output.

\descr{Canaries.} DP definitions assume a strong adversary that can in theory define the entire dataset given as input to a given mechanism (i.e., ``dataset canary'').
However, in the context of DP-SGD, auditing in this threat model typically entails training on so-called ``pathological'' datasets, which can destroy model utility~\cite{nasr2021adversary}  or may not reflect real-world threat models.
While these considerations do not affect the theoretical analysis of DP-SGD, which provides guarantees against the \emph{worst-case} adversary, they can inform of the privacy properties satisfied by the specific model being built.

Consequently, several DP-SGD auditing algorithms involve adversaries that can insert individual sample gradients at each step (i.e., ``gradient canary'')~\cite{nasr2023tight,annamalai2024what,cebere2025tighter}.
In practice, however, adversaries may only be able to insert a single sample into the input dataset (i.e., ``sample canary'').
While early work has focused on inserting dataset canaries~\cite{nasr2021adversary}, more recent research relies on inserting sample canaries instead, which is much more restrictive~\cite{pillutla2023unleashing,nasr2023tight,steinke2023privacy,annamalai2024nearly,mahloujifar2024auditing}.

To account for the use of canaries, in Algorithm~\ref{alg:audit_game}, the Adversary defines the entire dataset (line 1) only in the dataset canary threat model.
Whereas in the gradient and sample canary threat models, in line 1, the dataset $D^-$ is (possibly randomly) sampled from a ``natural'' distribution, e.g., CIFAR-10, instead.
Additionally, in the gradient canary model, the DP-SGD algorithm is modified to allow the Adversary to provide gradient canaries associated with the input canaries through the auxiliary information, which will be inserted by the algorithm in place of those computed on the input canaries.

\subsection{Attacks}
\label{sec:attack_defns}
When auditing DP implementations of simpler algorithms with low-dimensional output spaces (e.g., Laplace mechanism or Report Noisy Max)~\cite{ding2018detecting,bichsel2018dp,liu2019minimax,niu2022dp}, the DP auditing algorithm can directly analyze the outputs of the mechanism.
However, auditing more complex mechanisms (e.g., DP-SGD) with high-dimensional output spaces can be especially challenging, as it is unclear how to transform the high-dimensional output to a privacy parameter estimate. 
One popular way is to rely on privacy attacks to post-process the raw outputs from mechanisms into observations that can be thresholded and analyzed more easily.
We discuss these attacks and their role %
next.

\descr{Differential Privacy Distinguishability (DPD).} Here, we consider an adversary aiming to learn a ``distinguishing function'' to tell apart outputs from adjacent inputs~\cite{bichsel2021dp,bernau2021quantifying,lokna2023group,zhang2024dp,lu2024eureka,annamalai2024what,arcolezi2024revealing,askin2025general}. 
For instance, assume a machine learning model is trained on outputs from adjacent datasets $\mathcal{M}(D)$ and $\mathcal{M}(D')$ for binary classification (`0' for $\mathcal{M}(D)$ and `1' for $\mathcal{M}(D')$).
Then, \emph{fresh outputs} (not used in training) are sampled from the mechanism and converted into scalar ``scores'' by running inference on the model.
The scalar scores represent the adversary's confidence that the outputs came from processing $D'$ as opposed to $D$.

\descr{Membership Inference (MI).}
When auditing differentially private machine learning algorithms (e.g., DP-SGD, differentially private synthetic data), a common post-processing technique is membership inference (MI)~\cite{jagielski2020auditing,nasr2021adversary,lu2022general,houssiau2022tapas,tramer2022debugging,zanella2023bayesian,nasr2023tight,pillutla2023unleashing,steinke2023privacy,maddock2023canife,matsumoto2024measuring,chadha2024auditing,chida2024experimental,xiang2024preserving,annamalai2024what,debenedetti2024privacy,feng2024privacy,annamalai2024nearly,yoon2024optimizing,annamalai2024shuffle,annamalai2024s,nasr2025the,cebere2025tighter,xiang2025privacy,ganev2025elusive} owing to its popularity as a privacy measurement tool for (non-private) ML models.
In an MI attack, the adversary attempts to infer whether a given record was used to train the model.
This strongly aligns with the add/remove adjacency where either a canary is added to the dataset or not.

Typically, when auditing in the \texttt{Sample Canary} threat model, MI is run by computing the model's loss on the canary~\cite{jagielski2020auditing,nasr2021adversary,lu2022general,tramer2022debugging,zanella2023bayesian,nasr2023tight,pillutla2023unleashing,steinke2023privacy,annamalai2024what,feng2024privacy,annamalai2024nearly,mahloujifar2024auditing,yoon2024optimizing,annamalai2024s,nasr2025the} or by training shadow models similar the target model (in terms of model/training architecture and data distribution)~\cite{houssiau2022tapas,chida2024experimental,annamalai2024what,debenedetti2024privacy,ganev2025elusive}.
In the \texttt{Gradient Canary} threat model, the adversary can also compute the dot product~\cite{nasr2023tight,steinke2023privacy,maddock2023canife,xiang2024preserving,annamalai2024what,steinke2023privacy,mahloujifar2024auditing,xiang2025privacy} or cosine similarity~\cite{matsumoto2024measuring,galen2024oneshot,cebere2025tighter} between the gradient canary and noisy model updates or explicitly compute the likelihood ratio function~\cite{annamalai2024shuffle}.

\descr{Attribute Inference (AI).}
Although MI is the most common privacy attack for auditing DP-ML techniques, in specific circumstances, such as Label DP~\cite{ghazi2021deep} and synthetic data~\cite{annamalai2024what}, Attribute Inference (AI) has also been used~\cite{malek2021antipodes,houssiau2022tapas}.
In AI, the adversary attempts to infer specific private attributes of a given record, given the record's public attributes.
To perform auditing, Malek et al.~\cite{malek2021antipodes} infer the label associated with the canary sample in Label DP, while Houssiau et al.~\cite{houssiau2022tapas} infer sensitive attributes of a tabular canary given all other attributes for DP synthetic data.
Overall, establishing a baseline attack success rate is typically more complicated for AI~\cite{houssiau2022tapas}.

\descr{Reconstruction (Recon).}
Drawing on the theoretical bounds by Hayes et al.~\cite{hayes2023bounding}, reconstruction attacks (Recon) can also be used to audit DP implementations~\cite{mahloujifar2024auditing}.
Here, the adversary attempts to generate canaries from the data domain given only the output of the mechanism and the canary distribution.
Although this can be difficult for arbitrary canary distributions, Mahloujifar et al.~\cite{mahloujifar2024auditing} show that the problem is tractable when the canaries are uniformly sampled from discrete distributions each containing $k$ canaries.

\subsection{Evaluation Functions}
\label{sec:eval_fns_defns}

Once simple scalar scores have been extracted from the mechanism outputs, there are several ways to achieve the audit's goal.
While early work requires running the mechanism multiple times ($n_{\mathrm{R}} > 1$), to reduce computational overhead, recent work has focused on doing so in a single run ($n_{\mathrm{R}} = 1$).
In theory, the evaluation function can be specifically designed with the audit's goal in mind, i.e., estimation or verification.
In practice, however, evaluation functions are typically designed for the former, while the latter is done by simply checking if the estimated parameters are consistent with the claimed DP guarantees -- i.e., compute $(\empeps, \empdel) \leftarrow \zeta(\cdot)$ and check if $\empeps \leq \varepsilon$ and $\empdel \leq \delta$ for verification.
Next, we discuss estimation variants of different evaluation functions.

\subsubsection{Multiple Runs}

\descr{Output-set optimization.}
Early work on auditing DP implementations~\cite{ding2018detecting,bichsel2018dp,liu2019minimax,niu2022dp,lokna2023group} focused on simple algorithms (e.g., Laplace mechanism or Report Noisy Max), directly testing the approximate DP condition with respect to a set of outputs $E$, i.e., $\Pr[\mathcal{M}(D) \in E] \leq \mathrm{e}^\varepsilon \Pr[\mathcal{M}(D') \in E] + \delta$.
Since the outputs of these mechanisms are already simple scalars, the attack function $\mathcal{P}$ simply returns the output of the mechanism without modifying it.
To maximize the power of the audit, the evaluation function optimizes the output-set $E$ that results in the largest $\varepsilon$ for a given $\delta$.

\descr{Distribution estimation.}
Alternatively, one can empirically estimate the distributions the observations are drawn from.
For instance, Askin et al.~\cite{askin2022statistical} and Gorla et al.~\cite{gorla2023on} use kernel density estimators and histogram estimators, respectively, to estimate the distributions of observations drawn from $\mathcal{M}(D)$ and $\mathcal{M}(D')$.
After estimating, they test the approximate DP condition directly as done in output-set optimization.
Conversely, Kong et al.~\cite{kong2024dp} compute the Hockey-stick\footnote{$H_\alpha(P, Q) = \int_x \max\{P(x) - \alpha Q(x), 0\}$~\cite{sason2016f}.} or R{\'e}nyi divergence depending on the claimed DP guarantees, while Huang et al.~\cite{huang2025auditing} estimate the variance of the distributions and estimate the privacy guarantees from the variance.

\descrit{FPR/FNR.}
In the context of private ML, the most popular technique is to compute the false positive and false negative rates of the privacy attack and compare them with the bounds inferred from DP guarantees (cf.~Section~\ref{sec:hypo_testing_defn}).
First introduced by Jagielski et al.~\cite{jagielski2020auditing} for DP-SGD, this technique has been used in numerous research papers and extended to audit other DP mechanisms such as synthetic data generation, federated learning, and private prediction~\cite{nasr2021adversary,lu2022general,houssiau2022tapas,tramer2022debugging,zanella2023bayesian,nasr2023tight,maddock2023canife,matsumoto2024measuring,chadha2024auditing,chida2024experimental,xiang2024preserving,annamalai2024what,debenedetti2024privacy,feng2024privacy,annamalai2024nearly,yoon2024optimizing,annamalai2024shuffle,annamalai2024s,cebere2025tighter,nasr2025the,ganev2025elusive}.

One drawback is that this requires thousands, if not millions, of observations to provide reasonable confidence intervals~\cite{annamalai2024shuffle}.
Lu et al.~\cite{lu2022general} and Zanella-Beguelin et al.~\cite{zanella2023bayesian} attempt to reduce the number of runs required by optimizing the confidence interval computation but still require around a thousand runs to audit.
Nasr et al.~\cite{nasr2023tight} reduce the number of runs to two by auditing each step of the DP-SGD algorithm independently.
Similarly, Maddock et al.~\cite{maddock2023canife} only require a single run of the DP-FedSGD~\cite{mcmahan2018learning} algorithm, although they audit each round independently.%

\descr{Multiple Canaries.} %
While previous work~\cite{ding2018detecting,jagielski2020auditing,nasr2021adversary,nasr2023tight} has focused on techniques that add only a single canary in each run, Pillutla et al.~\cite{pillutla2023unleashing} use multiple canaries per run.
Specifically, they perform multiple hypothesis tests (one for each canary) for each mechanism output, thus reducing the number of runs required to a few hundred.
However, for computationally expensive algorithms like DP-SGD, even a few hundred runs can make it computationally prohibitive to audit in practice, especially for large language models~\cite{panda2025privacy}.

\subsubsection{One Run}

\descr{Guessing Game.}
As mentioned, recent work has focused on reducing the number of runs needed to audit mechanisms to only one.
This is formalized through a {\em guessing game} where DP guarantees are estimated using the \emph{accuracy} of a privacy attack instead of FPR/FNRs.
Malek et al.~\cite{malek2021antipodes} introduce a heuristic to audit DP mechanisms by inserting multiple canaries into the mechanisms' input.
Steinke et al.~\cite{steinke2023privacy} formalize this heuristic for MI and demonstrate that MI attacks targeting multiple canaries can indeed be used for DP auditing, albeit only achieving loose audits.

Mahloujifar et al.~\cite{mahloujifar2024auditing} and Xiang et al.~\cite{xiang2025privacy} improve by auditing with, respectively, $f$-DP and information theory.
Specifically, Mahloujifar et al.~\cite{mahloujifar2024auditing} extend~\cite{steinke2023privacy} to cover $f$-DP, audit using Recon attacks, and improve the bound on attack success of Hayes et al.~\cite{hayes2023bounding}. %
Xiang et al.~\cite{xiang2025privacy} stick to MI but model the auditing task using information-theoretic principles (i.e., noisy channels). %
Beyond fully connected neural networks and convolutional neural networks, recent work by Panda et al.~\cite{panda2025privacy} uses the techniques from~\cite{steinke2023privacy} to optimize canary design and audit large language models.

\descr{Variance estimation.}
In the context of Federated Learning, Galen et al.~\cite{galen2024oneshot} audit the DP-FedAvg algorithm by inserting multiple independent isotropically distributed canaries into each round of the algorithm.
They then estimate the mean and variance of the cosine similarities between the inserted canaries and the mechanism's output.

\subsection{Confidence Level}
\label{sec:conf_lev_defn}
Whether the audit goal is verification or estimation, as mentioned in Section~\ref{sec:goals_of_auditing}, performing it at a given confidence level can be important to provide auditors with statistical assurance that the result of the audit is correct.

One popular method is to estimate the type I and type II errors of the privacy attack to a given level of confidence using statistical tools like the Clopper-Pearson~\cite{clopper1934use} interval~\cite{jagielski2020auditing,bichsel2021dp,nasr2021adversary,malek2021antipodes,houssiau2022tapas,niu2022dp,tramer2022debugging,lokna2023group,zanella2023bayesian,nasr2023tight,maddock2023canife,matsumoto2024measuring,chadha2024auditing,chida2024experimental,annamalai2024what,annamalai2024nearly,yoon2024optimizing,annamalai2024shuffle,annamalai2024s,cebere2025tighter,nasr2025the,arcolezi2024revealing,ganev2025elusive}, Katz-log~\cite{katz1978obtaining} interval~\cite{lu2022general,nasr2023tight}, or credible intervals~\cite{zanella2023bayesian,nasr2023tight,xiang2024preserving,debenedetti2024privacy,galen2024oneshot}.
Some audits also use custom methods to specify the confidence level tuned to their specific auditing algorithms~\cite{ding2018detecting,bichsel2021dp,bernau2021quantifying,askin2022statistical,pillutla2023unleashing,steinke2023privacy,gorla2023on,kong2024dp,mahloujifar2024auditing,xiang2025privacy,askin2025general}.
However, some do not report any confidence intervals at all~\cite{bichsel2018dp,liu2019minimax,wang2021current,zhang2024dp,lu2024eureka,huang2025auditing}.
\subsection{Remarks}

Overall, while DP audits have used a wide range of techniques, very few audits have achieved all three desired properties, and only for simple DP mechanisms.

\section{Efficient, End-to-End, and Tight DP Audits?}
\label{sec:findings}
Based on our systematic review of the DP auditing body of work, we now discuss several key details of DP auditing that have been neglected by prior work, discover limiting factors to achieving all three desiderata, and identify novel directions for future work to focus on.

\subsection{DP Guarantees \& Adjacency}
\label{sec:dp_defns_analysis}
Naturally, the DP guarantee variant being audited is a fundamental requirement of any DP auditing algorithm, as per Definition~\ref{def:dp_audit_alg}.
As evident from Table~\ref{tab:summary_intro}, while earlier work~\cite{ding2018detecting,bichsel2021dp,jagielski2020auditing,nasr2021adversary,houssiau2022tapas,askin2022statistical,lokna2023group} focuses on simpler variants like \puredp and \approxdp, more recent audits~\cite{nasr2023tight,annamalai2024nearly,mahloujifar2024auditing,cebere2025tighter,xiang2025privacy} do so on \fdp and \gdp to achieve tight audits. %

This is because simpler variants of DP (\puredp and \approxdp) may not always describe the privacy leakage of specific mechanisms optimally, i.e., the \emph{actual} privacy leakage of the mechanism may be far from the \emph{theoretical} privacy leakage expected from simple DP guarantees.
In turn, audits are loose since DP auditing measures the actual privacy leakage from mechanisms.
For instance, Nasr et al.~\cite{nasr2023tight} achieve tight audits of DP-SGD using the \fdp guarantee, which they show models the actual privacy leakage of DP-SGD much more closely compared to \approxdp initially used for auditing~\cite{jagielski2020auditing,nasr2021adversary}.
Therefore, recent audits of DP-SGD use the \fdp or the related \gdp guarantee instead~\cite{nasr2023tight,annamalai2024nearly,mahloujifar2024auditing,cebere2025tighter,xiang2025privacy}.

On the other hand, auditing with \fdp poses several challenges.
Firstly, as discussed in Section~\ref{sec:subsampling_analysis}, \fdp guarantees of mechanisms are not always known (this is an active area of research).
Furthermore, when used for DP estimation, it can be difficult to estimate the full $f$-DP curve compared to DP guarantees with only a few scalar parameters (e.g., \puredp, \approxdp, \gdp).
In fact, even when estimating \approxdp, the $\delta$ parameter is typically fixed~\cite{jagielski2020auditing,nasr2021adversary,houssiau2022tapas,askin2022statistical,lokna2023group,pillutla2023unleashing,steinke2023privacy,annamalai2024what,feng2024privacy,annamalai2024shuffle} so that only $\varepsilon$ is being estimated.
Therefore, there are no known methods to estimate $f$-DP guarantees with confidence intervals, and doing so constitutes an open research question.

Moreover, the DP guarantees and adjacency notions audited might not always be explicitly stated or are left unclear.
This is the case for the adjacency notion in several key papers~\cite{feng2024privacy,annamalai2024nearly,huang2025auditing,nasr2025the,askin2025general}.
Furthermore, several results~\cite{nasr2023tight,annamalai2024nearly,mahloujifar2024auditing,cebere2025tighter,xiang2025privacy} are given with respect to \approxdp, although closer inspection of their methodology reveals that they are instead auditing the \fdp or \gdp guarantees.

\begin{takeawaybox}{Takeaway 1}
    DP audits should be clear about the exact DP guarantee being audited along with the adjacency notions considered.
    In general, auditing \fdp guarantees are optimal, but deriving tight guarantees for specific mechanisms is currently an open challenge.
\end{takeawaybox}

\subsection{Sub-sampling Schemes}
\label{sec:subsampling_analysis}

Another key detail in DP auditing is the underlying sub-sampling scheme.
From Table~\ref{tab:summary_intro}, we observe that the most common sub-sampling scheme audited is Poisson~\cite{jagielski2020auditing,nasr2021adversary,nasr2023tight,maddock2023canife,pillutla2023unleashing,steinke2023privacy,annamalai2024what,feng2024privacy,mahloujifar2024auditing,cebere2025tighter,xiang2025privacy}, owing to its use in the DP-SGD algorithm.
However, recent work~\cite{galen2024oneshot,annamalai2024shuffle} has also looked at auditing the shuffling scheme.
Overall, while audits not involving sub-sampling schemes are known to be tight~\cite{ding2018detecting,bichsel2021dp,askin2022statistical,lokna2023group,nasr2023tight,mahloujifar2024auditing,xiang2025privacy}, tight audits involving sub-sampling schemes are much rarer~\cite{nasr2021adversary,annamalai2024what,feng2024privacy}, usually requiring additional assumptions, such as step-by-step auditing or simple models.

In this context, the main challenge %
is having tight DP guarantees of mechanisms employing sub-sampling schemes, i.e., knowing that the \emph{actual} privacy leakage of mechanisms matches the \emph{theoretical} privacy leakage expected from the DP guarantee.
Next, we focus on the Gaussian Mechanism since sub-sampling schemes are most commonly used in conjunction with that, e.g., in DP-SGD~\cite{abadi2016deep}.
In Table~\ref{tab:subsampling}, we summarize the sub-sampling schemes and adjacency notions for which tight DP guarantees are currently known.
While tight guarantees are known for Poisson sub-sampling for any adjacency and Sampling WOR, Sampling WR, and Balls-in-Bins sampling for the Add/Remove, Edit, and Zero-out adjacencies, respectively,
for the vast majority of sub-sampling scheme and adjacency pairs, tight guarantees remain unknown.

\begin{takeawaybox}{Takeaway 2}
    Deriving the tight guarantees for many popular sub-sampling schemes is an open research question; thus, the ability to tightly audit mechanisms that use them is currently limited.
\end{takeawaybox}

\begin{table}[t]
\small
		\setlength{\tabcolsep}{2pt}
    \begin{tabular}{lccccc}
\toprule
 & \textbf{Poisson} & \textbf{Sampling} & \textbf{Sampling} & \textbf{Shuffling} & \textbf{Balls-in-} \\
 & & \edit{{\bf WR}} & \edit{{\bf WOR}} && {\bf Bins} \\
        \midrule
\textbf{Add/Remove} & ~\cite{zhu2022optimal} \cellcolor{pos} & \cellcolor{neg} & ~\cite{lebeda2025avoiding} \cellcolor{pos} & \cellcolor{neg} & \cellcolor{neg} \\
\textbf{Edit} & \cellcolor{pos}~\cite{koskela2020computing,lebeda2025avoiding} & ~\cite{schuchardt2024unified} \cellcolor{pos} & ~\cite{zhu2022optimal,lebeda2025avoiding} \cellcolor{semipos} & \cellcolor{neg} & \cellcolor{neg} \\
\textbf{Zero-Out} & ~\cite{chua2024private} \cellcolor{pos} & \cellcolor{neg} & \cellcolor{neg} & ~\cite{chua2024private,chua2024scalable}\cellcolor{semipos} & ~\cite{chua2025ballsandbins} \cellcolor{pos} \\ \bottomrule
    \end{tabular}
    \centering
    \caption{Summary of whether tight guarantees of the Gaussian Mechanism are currently known for sub-sampling schemes and adjacency notions. NB: Cells are colored green if prior work has identified tight guarantees, yellow if tight guarantees are not currently known, and brown otherwise.}
    \label{tab:subsampling}
\end{table}

\subsection{Threat Models}
\label{sec:threat_model_analysis}
One elusive combination of desired properties for audits of complex mechanisms like DP-SGD is for them to be both end-to-end and tight.
Specifically, Table~\ref{tab:summary_intro} reveals that end-to-end audits of DP-SGD are not tight~\cite{jagielski2020auditing,nasr2021adversary,nasr2023tight,pillutla2023unleashing,steinke2023privacy,annamalai2024nearly,mahloujifar2024auditing,cebere2025tighter} and viceversa~\cite{nasr2021adversary,nasr2023tight}.
The only exception is the work by Feng et al.~\cite{feng2024privacy}, who achieve both tight and end-to-end audits of DP-SGD, but only for simple logistic regression models.

\begin{table}[t]
\newcolumntype{P}[1]{>{\centering\arraybackslash}p{#1}} %
\small
		\setlength{\tabcolsep}{4pt}

    \begin{tabular}{lP{2.2cm}P{2cm}c}
\toprule
         & \textbf{Sample} & \multicolumn{1}{c}{\textbf{Gradient}} & \multicolumn{1}{c}{\textbf{Dataset}} \\
        \midrule
        {\bf Intermediate} & ~\cite{huang2025auditing} \cellcolor{neg} & ~\cite{bernau2021quantifying,nasr2021adversary,nasr2023tight,steinke2023privacy,pillutla2023unleashing,matsumoto2024measuring,mahloujifar2024auditing,xiang2025privacy} \cellcolor{pos} & ~\cite{nasr2021adversary} \cellcolor{pos} \\
        \textbf{Final} & ~\cite{jagielski2020auditing,nasr2021adversary,lu2022general,zanella2023bayesian,nasr2023tight,steinke2023privacy,debenedetti2024privacy,feng2024privacy,mahloujifar2024auditing,nasr2025the,xiang2025privacy,panda2025privacy} \cellcolor{neg} & ~\cite{cebere2025tighter} \cellcolor{neg} & ~\cite{nasr2021adversary,annamalai2024s,askin2025general} \cellcolor{pos} \\ \bottomrule
    \end{tabular}
    \centering
    \caption{Summary of different threat models -- specifically, visibility of intermediate vs.~final models -- for DP-SGD auditing considered in prior work for different canary types. NB: Cell colored green if tight auditing has been achieved  and brown otherwise.}
    \label{tab:threat_model}
\end{table}

To analyze this further, we break down end-to-end audits into specific adversarial capabilities (model visibility and canary type) as explained in Section~\ref{sec:threat_model_defn}.
In Table~\ref{tab:threat_model}, we highlight the threat models where state-of-the-art DP-SGD audits are tight.
Generally, tight auditing in stronger threat models has been accomplished; however, while there has been significant work on the weakest \texttt{(Final Model, Sample Canary)} (aka ``Black-box'') threat model, less work has focused on the slightly stronger \texttt{(Final Model, Gradient Canary)} and \texttt{(Intermediate Models, Sample Canary)} models.\footnote{To ease presentation, we use the \texttt{(Model Visibility, Canary Type)} notation to quickly refer to the corresponding threat models.}

Overall, we consider a DP audit ``end-to-end'' only in the weakest threat model \texttt{(Final Model, Sample Canary)}.
Given that these audits are still far from tight, focusing efforts on the two slightly stronger threat models would be promising interim directions.

\begin{takeawaybox}{Takeaway 3}
    More research on auditing DP-SGD in the \texttt{\justify (Final Model, Gradient Canary)} and \texttt{\justify (Intermediate Models, Sample Canary)} threat models can potentially pave the way toward tight audits in \texttt{\justify (Final Model, Sample Canary)}, which is crucial for end-to-end audits.
\end{takeawaybox}

\subsection{Evaluation Functions}
Another elusive combination of the desiderata is accomplishing DP audits that are both efficient and tight.
From Table~\ref{tab:summary_intro}, we observe that only recently this has been achieved, albeit only for simple mechanisms like 
 Gaussian~\cite{mahloujifar2024auditing,xiang2025privacy}.
As a result, an important problem for future research is designing evaluation functions that only require mechanisms to be run once for auditing. %

As discussed in Section~\ref{sec:eval_fns_defns}, Nasr et al.~\cite{nasr2023tight}'s techniques only require two runs but they only manage to audit a single step of the DP-SGD mechanism.
Steinke et al.~\cite{steinke2023privacy}, Mahloujifar et al.~\cite{mahloujifar2024auditing}, Panda et al.~\cite{panda2025privacy}, and Xiang et al.~\cite{xiang2025privacy} all propose techniques to audit the entire DP-SGD mechanism in one run.
However, even in the strong \texttt{(Gradient Canary, Intermediate Models)} threat model, their audits remain loose, at best only achieving an empirical lower bound of $\empeps \approx 4.5$ for a theoretical $\varepsilon = 8$~\cite{xiang2025privacy}.
Similarly, one-run audits of Federated Learning variants of DP-SGD (i.e., DP-FedAvg and DP-FedSGD) are also loose~\cite{maddock2023canife,galen2024oneshot}.
By contrast, tight one-run audits for simpler mechanisms such as the Gaussian Mechanism are possible~\cite{xiang2025privacy}.
This suggests there may be fundamental challenges that may need to be overcome when auditing DP-SGD in one run---we discuss this in more detail in Section~\ref{sec:open_rqs}.

\begin{takeawaybox}{Takeaway 4}
    Efficient auditing in a single run is a primary focus of recent DP auditing work but, even in powerful threat models, e.g., (\texttt{Gradient Canary}, \texttt{Intermediate Models}), audits remain loose.
    Given the complications surrounding auditing in weaker threat models (see Section~\ref{sec:threat_model_analysis}), research should focus on achieving tight audits in one run in these more powerful threat models first.
\end{takeawaybox}

\subsection{Privacy Region \& Confidence Level}\label{sec:region}

\captionsetup[subfigure]{justification=centering}
\begin{figure*}[t]
    \centering
    \begin{subfigure}[t]{0.495\textwidth}
    \centering
        \includegraphics[width=0.55\linewidth]{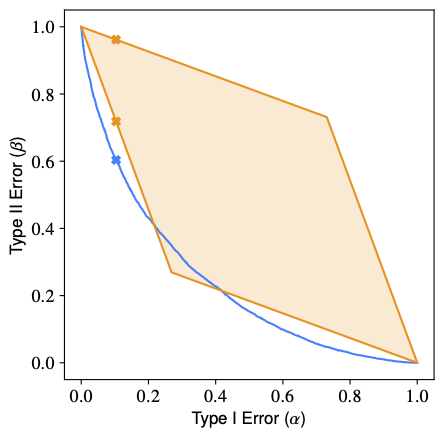}
        \vspace{-0.2cm}
        \caption{Verifying (incorrectly) claimed $(1, 10^{-5})$-DP guarantees.}
        \label{fig:pr_verify}
    \end{subfigure}
    \begin{subfigure}[t]{0.495\textwidth}
    \centering
        \includegraphics[width=0.6\linewidth]{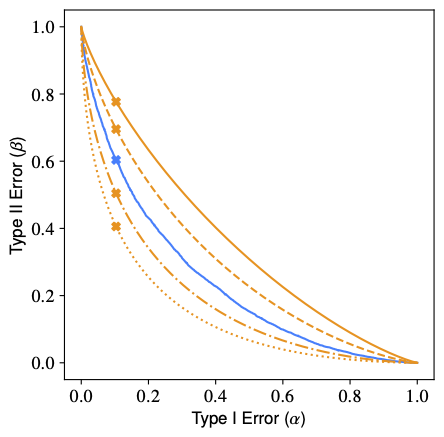}
        \vspace{-0.2cm}
        \caption{Estimating empirical privacy guarantees using \gdp, with $\mu = \{0.5, 0.75, 1.25, 1.5\}$.}
        \label{fig:pr_estimate}
    \end{subfigure}
    \caption{Privacy regions when auditing Gaussian mechanism ($\sigma = 1$) with \approxdp and \gdp. The blue line indicates estimated guarantees and the orange lines (and shaded regions) represent claimed or candidate guarantees.}
    \label{fig:pr_audit}
\end{figure*}

The concept of \emph{privacy region} has been used by auditors to both verify and estimate DP guarantees of the mechanisms under study.
Figure~\ref{fig:pr_audit} visualizes how the Gaussian mechanism ($\sigma = 1$) gets audited under both the \approxdp and \gdp variants.
More precisely, in Figure~\ref{fig:pr_verify}, the privacy region is estimated from the mechanism and compared against the hypothetical guarantees of $(1, 10^{-5})$-DP, which the mechanism does \emph{not} satisfy.
Note the presence of large sections of the estimated privacy region that fall well outside the shaded theoretical privacy region for $(1, 10^{-5})$-DP.
This suggests that the claimed DP guarantees are wrong and they can therefore be refuted experimentally.

On the other hand, in Figure~\ref{fig:pr_estimate}, the estimated privacy region is compared against multiple ``candidate'' \gdp regions ($\mu = 0.5, 0.75, 1.25, 1.5$) to estimate the empirical $\mu_{\mathrm{emp}}$-GDP guarantees of the mechanism.
The $\mu$ parameter is changed to plot different theoretical privacy regions (for simplicity, we only plot the lower boundaries).
By doing so, we can estimate that the mechanism satisfies $\mu_{\mathrm{emp}}$-GDP for $0.75 \leq \mu_{\mathrm{emp}} \leq 1.25$, which we can further fine-tune by comparing \gdp regions at lower resolutions.

In both cases, either the full privacy region can be estimated (represented by the line), or a single point estimate can be used (represented by the markers).
This choice mainly affects the \emph{confidence level} of the overall audit procedure as confidence levels derived for single-point estimates may not necessarily transfer to full-region estimates.

DP auditing research typically bypasses computing valid confidence intervals to ease experimentation.
As mentioned in Section~\ref{sec:conf_lev_defn}, some researchers either leave that out altogether~\cite{bichsel2018dp,liu2019minimax,wang2021current,zhang2024dp,lu2024eureka,huang2025auditing} or they estimate the full privacy region but report confidence intervals for a single-point estimate instead~\cite{nasr2021adversary,maddock2023canife,nasr2023tight,annamalai2024nearly}, thus making them invalid.
Nevertheless, in practice, audits should ensure that valid confidence intervals are computed for the auditing method used.
For instance, the best possible single-point estimate can be derived using ``shadow models''~\cite{annamalai2024what} or the entire region can be estimated using \gdp~\cite{nasr2023tight}.
Additionally, further research could also focus on reliably estimating full privacy regions with meaningful confidence intervals.

\begin{takeawaybox}{Takeaway 5}
    When deploying auditing algorithms in the real world, care has to be taken to ensure that the confidence intervals computed are \emph{valid}.
    Further research could also focus on doing so when estimating full privacy regions.
\end{takeawaybox}

\section{Discussion \& Conclusion}

This paper systematized research on DP auditing using a systematic framework geared to categorize the underlying settings/techniques and measure progress in the state of the art along three key desired properties: \textit{efficiency}, \textit{end-to-end-ness}, and \textit{tightness}.
Overall, our analysis showed that while DP auditing made significant and rapid progress in making audits more effective, useful, and easily deployable on production systems, there are still areas with substantial research gaps.
In the process, we identified several key insights and open problems; we are confident that future work can build upon these to establish a ``Gold Standard'' for DP auditing.

In the rest of this section, we %
present actionable recommendations, and highlight open research questions to prioritize.

\subsection{What is considered SOTA in DP Auditing?}
\label{sec:sota}

Although there are different settings and thus different auditing techniques in the literature, we attempt to identify the results that can be considered state-of-the-art (SOTA) with respect to tightness.
To do so, we compare the reported empirical $\varepsilon$ values at standardized theoretical $\varepsilon$ levels (i.e., $\varepsilon = 8$) and algorithms (i.e., training CNN on CIFAR-10~\cite{jagielski2020auditing,nasr2021adversary,nasr2023tight,steinke2023privacy,annamalai2024nearly,cebere2025tighter,mahloujifar2024auditing,xiang2025privacy} when auditing DP-SGD).
More precisely, in Table~\ref{tab:sota}, we do so while breaking down the DP guarantees audited and whether the other two key desirable properties are achieved.

Put simply, when efficient audits are required, Xiang et al.~\cite{xiang2025privacy}'s work is SOTA.
Specifically, their techniques achieve tight audits for simple mechanisms such as Laplace, Gaussian, or Randomized Response but only loose audits for more complex mechanisms like DP-SGD. 
One main limitation of Xiang et al.~\cite{xiang2025privacy}'s work is the potential loss in model utility as it requires millions of target records to be added to the initial dataset for audits to be tight.
For DP-SGD specifically, loosening the end-to-end property allows us to achieve tight audits using Nasr et al.~\cite{nasr2023tight}'s technique, while also preserving model utility.

When audits can be inefficient but minimal utility loss is expected, we consider the method by Jagielski et al.~\cite{jagielski2020auditing} to be SOTA for complex algorithms like DP-SGD and Lokna et al.~\cite{lokna2023group} for simple mechanisms satisfying \approxdp.
Also, Chadha et al.~\cite{chadha2024auditing}'s method is the {\em only} technique available for mechanisms satisfying \rdp.
Finally, for \fdp (or \gdp), Annamalai et al.~\cite{annamalai2024nearly}'s auditing method is SOTA for end-to-end audits, while that by Cebere et al.~\cite{cebere2025tighter} is SOTA for audits over sub-sampling that do not need to be end-to-end.

\begin{table}[t]
\small
		\setlength{\tabcolsep}{4pt}
    \begin{tabular}{l@{}c@{}cc}
        \toprule
        {\bf DP Guarantee} & {\bf End-to-End} & {\bf Efficient} & {\bf Not Efficient} \\
        \midrule
        \multirow{2}{*}{\approxdp, \puredp} & \cmark & \multirow{2}{*}{Xiang et al.~\cite{xiang2025privacy}} & Jagielski et al.~\cite{jagielski2020auditing}, \\
        & \cmark & &  Lokna et al.~\cite{lokna2023group}\\
        \midrule
        \rdp & \xmark & -- & Chadha et al.~\cite{chadha2024auditing}\\
        \midrule
        \multirow{2}{*}{\fdp, \gdp} & \xmark & Nasr et al.~\cite{nasr2023tight} & Cebere et al.~\cite{cebere2025tighter} \\
         & \cmark & Xiang et al.~\cite{xiang2025privacy} & Annamalai et al.~\cite{annamalai2024nearly}  \\
        \bottomrule
    \end{tabular}
    \small
    \centering
    \caption{DP auditing methods considered to be state-of-the-art (SOTA) with respect to tightness, broken down by DP guarantee, efficiency, and end-to-end-ness. %
    (To ease presentation, we only report the last name of each paper's first author.)}
    \label{tab:sota}
\end{table}

\subsection{Recommendations}

\noindent{\bf Auditing Details.} As discussed in Section~\ref{sec:dp_defns_analysis}, one not-so-uncommon limitation in DP auditing research is that key details of the auditing methodology are spread across various sections of the paper and may not be explicitly stated or left unclear.
We suggest that researchers explicitly state the DP guarantee audited, along with the adjacency notion, sub-sampling scheme, threat model, and confidence interval considered.

\begin{figure}[t]
  \centering
  \includegraphics[width=0.75\linewidth]{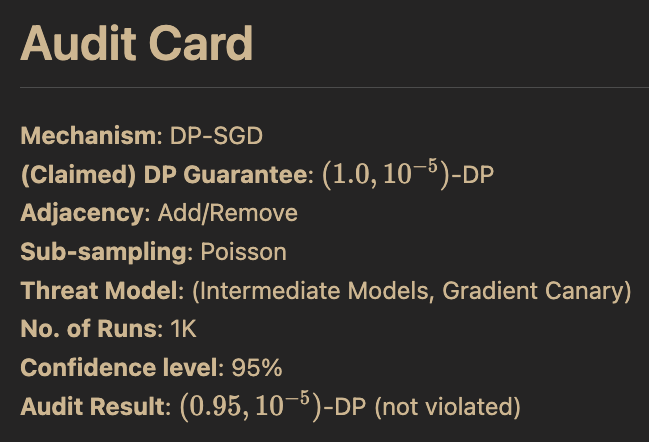}
  \caption{Example of a Possible Design for Audit Cards.}
  \label{fig:audit_card}
\end{figure}

For instance, they could adopt %
an ``Auditing Card'' to be added in the appendix and/or in code repositories, similar to the ``Model Cards'' released for models published on HuggingFace.\footnote{See \url{https://huggingface.co/docs/hub/en/model-cards}.}
We include a draft template in Figure~\ref{fig:audit_card}; future work could conduct a user study with researchers and practitioners working on DP auditing and explore the viability %
of this approach.

\descr{Tighter \fdp or \gdp guarantees.} As also discussed by Gomez et al.~\cite{gomez2025varepsilon}, we advocate for the standardized release and auditing of tighter \fdp or \gdp guarantees.
For example, when auditing the Gaussian mechanism, future audits could estimate and compare the empirical $\mu_{\mathrm{emp}}$ value to the theoretical \gdp guarantee, instead of converting both to a \approxdp value with an arbitrarily chosen $\delta$.
For mechanisms without a \gdp guarantee, the full \fdp curve could be estimated, and the $\Delta^{\leftrightarrow}$-divergence~\cite{kaissis2024beyond,gomez2025varepsilon} between the estimated and theoretical curve could be used as a ``goodness of fit'' metric.

\subsection{Open Research Questions}
\label{sec:open_rqs}
In conclusion, our systematization effort highlights the presence of at least four open research questions.

\descr{RQ1: Can we derive tight guarantees for all sub-sampling schemes?}
Although tight theoretical guarantees have been derived for many popular sub-sampling schemes (e.g., Poisson), much less is known about schemes like sampling without replacement and shuffling.
These schemes are commonly used to train models efficiently~\cite{annamalai2024shuffle}, and in the context of auditing, having tight guarantees is necessary for audits to be tight.
Thus, deriving tight theoretical guarantees for these schemes remains an important open problem.

\descr{RQ2: Are end-to-end and tight audits possible?}
One of the most elusive combinations of desired properties is end-to-end-ness and tightness.
While this is possible for simple mechanisms, for complex algorithms like DP-SGD, very little work has come close. %
Based on our systematic analysis, we believe this question can be further broken down into investigating slightly weaker threat models, namely, \texttt{(Final Model, Gradient Canary)} and \texttt{(Intermediate Models, Sample Canary)} before addressing the end-to-end threat model, i.e., (\texttt{Final Model, Sample Canary}).

Cebere et al.~\cite{cebere2025tighter} audit DP-SGD in \texttt{(Final Model, Gradient Canary)} but is not tight, suggesting that tighter theoretical analysis might be necessary, which is in itself an open research question~\cite{annamalai2024s,nasr2025the}.
Hence, we call for future work to focus on the \texttt{(Final Model, Gradient Canary)} and \texttt{(In\-ter\-me\-di\-ate} \texttt{Models, Sample Canary)} threat models. %

\descr{RQ3: Are efficient and tight audits of DP-SGD possible?}
Another elusive combination of properties is efficiency and tightness.
While several methods have been suggested toward this goal~\cite{steinke2023privacy,mahloujifar2024auditing,xiang2025privacy}, only recently Xiang et al.~\cite{xiang2025privacy} present efficient and tight audits, albeit only for simple mechanisms.
However, while recent work has identified fundamental challenges in efficiently auditing complex mechanisms~\cite{keinan2025well}, these challenges do not particularly limit DP-SGD; thus, efficient and tight audits of DP-SGD could be possible. 
Similar to RQ2, it might be useful to first consider alternative threat models to elicit the maximal privacy leakage in efficient audits.

\descr{RQ4: How do we make audits more \emph{robust}?} 
Although we have mainly focused on the three desired properties mentioned previously, recent research~\cite{wang2024curator} has suggested that other properties might be important to achieve in DP auditing as well.
Specifically, Wang et al.~\cite{wang2024curator} find that in some cases, it may be possible for malicious actors to pass DP auditing techniques with overstated privacy guarantees, reducing the reliability and robustness of audit techniques.

On the other hand, even though theoretical DP guarantees are all-encompassing, holding for all adversaries and inputs, DP audits are fundamentally empirical and can only test a finite subset of them. 
As a result, current DP auditing techniques only focus on a single class of attacks, which makes it challenging for them to identify bugs that are only observable under specific classes of attacks.
This also reduces the robustness of current DP auditing techniques and makes them brittle in practice.
Overall, we find that the robustness of DP audits is another important property of DP auditing that is currently understudied and deserving of the attention of privacy researchers.

\descr{Acknowledgements.}
This work has been supported by the National Science Scholarship (PhD) from the Agency for Science, Technology and Research, Singapore.

{\small
\bibliographystyle{abbrv}

}

\appendix

\section{Scopus Search Query}
\label{sec:exact_search}
Our search query was set up to catch the following keywords in the title or abstract of the papers: \textit{differential privacy}, \textit{differentially private}, \textit{audit}, \textit{find/detect/discover privacy violations/vulnerabilities}, \textit{break/violate DP}, \textit{estimate privacy}.
The exact search query we ran on the Scopus research database in January 2025 to collate the list of relevant prior work is provided in Figure~\ref{fig:query}.

\begin{figure}[h]
\begin{framed}
\begin{quote}
\footnotesize
\texttt{(\\
\-\hspace{0.5em} TITLE-ABS-KEY("differential privacy") OR \\
\-\hspace{0.5em} TITLE-ABS-KEY("differentially private") \\
) AND \\
(\\
\-\hspace{0.5em} TITLE-ABS-KEY("audit*") OR \\
\-\hspace{0.5em} TITLE-ABS-KEY("find* violat*") OR \\
\-\hspace{0.5em} TITLE-ABS-KEY("detect* violat*") OR \\
\-\hspace{0.5em} TITLE-ABS-KEY("identify* violat*") OR \\
\-\hspace{0.5em} TITLE-ABS-KEY("discover* violat*") OR \\
\-\hspace{0.5em} TITLE-ABS-KEY("find* vuln*") OR \\
\-\hspace{0.5em} TITLE-ABS-KEY("detect* vuln*") OR \\
\-\hspace{0.5em} TITLE-ABS-KEY("identify* vuln*") OR \\
\-\hspace{0.5em} TITLE-ABS-KEY("discover* vuln*") OR \\
\-\hspace{0.5em} TITLE-ABS-KEY("find* privacy violat*") OR \\
\-\hspace{0.5em} TITLE-ABS-KEY("detect* privacy violat*") OR \\
\-\hspace{0.5em} TITLE-ABS-KEY("identify* privacy violat*") OR \\
\-\hspace{0.5em} TITLE-ABS-KEY("discover* privacy violat*") OR \\
\-\hspace{0.5em} TITLE-ABS-KEY("find* privacy vuln*") OR \\
\-\hspace{0.5em} TITLE-ABS-KEY("detect* privacy vuln*") OR \\
\-\hspace{0.5em} TITLE-ABS-KEY("identify* privacy vuln*") OR \\
\-\hspace{0.5em} TITLE-ABS-KEY("discover* privacy vuln*") OR \\
\-\hspace{0.5em} TITLE-ABS-KEY("violat* DP") OR \\
\-\hspace{0.5em} TITLE-ABS-KEY("break* DP") OR \\
\-\hspace{0.5em} TITLE-ABS-KEY("privacy estimat*") OR \\
\-\hspace{0.5em} TITLE-ABS-KEY("estimat* privacy")\\
)\\[-5ex]}
\end{quote}
\end{framed}
\vspace{-8pt}
\caption{Scopus Query.}
\label{fig:query}
\end{figure}

\newpage
\section{Prior Work}
\label{sec:full_summary}
In Table~\ref{tab:summary_full}, we provide the full list of the 45 papers that form the basis of our systematization.
Whereas, as discussed in Section~\ref{sec:body_of_work}, Table~\ref{tab:summary_intro} only reports a selection of 21 %
papers (specifically, papers that are the first to audit a particular DP guarantee or mechanism, or that achieve the current ``best'' auditing results).

\onecolumn
\begin{footnotesize}
		\setlength{\tabcolsep}{3pt}
    \begin{longtable}{@{}l@{}cc@{}cc@{}cc@{}c@{}c}
        \toprule
        & & \multicolumn{2}{c}{{\bf Foundations}} & \multicolumn{2}{c}{{\bf Operational Details}} & \multicolumn{3}{c}{{\bf Progress}} \\
        \cmidrule(lr){3-4}        \cmidrule(lr){5-6}         \cmidrule(lr){7-9}
        {\bf Reference} & \makecell{{\bf Mechanisms}\\{\bf Audited}} & {\bf DP Guarantee~} & {\bf Sub-Sampling} & {\bf Attack} & \makecell{{\bf Eval.}\\{\bf Function}} & {\bf Efficient~} & {\bf End-to-End~} & {\bf Tight} \\
        \midrule
        Ding et al.~(2018)~\cite{ding2018detecting} & \makecell{Report Noisy Max\\Noisy Hist\\SVT} & \puredp & -- & -- & Output-set & \xmark & \cmark & \cmark \\
        \myrule
        Bischel et al.~(2018)~\cite{bichsel2018dp} & \makecell{Report Noisy Max\\NoisySum\\AboveThreshold} & \puredp & -- & -- & Output-set & \xmark & \cmark & \cmark \\
        \myrule
        Liu et al.~(2019)~\cite{liu2019minimax} & \makecell{Report Noisy Max\\Noisy Hist\\SVT\\Truncated Geometric} & \approxdp & -- & -- & Output-set & \xmark & \cmark & \cmark \\
        \myrule
        Jagielski et al.~(2020)~\cite{jagielski2020auditing} & DP-SGD & \approxdp & Poisson & MI & FPR/FNR & \xmark & \cmark & \xmark \\
        \myrule
        Bischel et al.~(2021)~\cite{bichsel2021dp} & \makecell{Report Noisy Max\\Noisy Hist\\SVT\\RAPPOR\\Prefix Sum\\Truncated Geometric} & \approxdp & -- & DPD & Output-set & \xmark & \cmark & \cmark \\
        \myrule
        Bernau et al.~(2021)~\cite{bernau2021quantifying} & DP-SGD & \approxdp & Poisson & \makecell{DPD\\MI} & Output-set & \xmark & \xmark & \cmark \\
        \myrule
        Wang et al.~(2021)~\cite{wang2021current} & \makecell{Laplace} & \puredp & -- & DPD & Output-set & \xmark & \cmark & \cmark \\
        \myrule
        \myspan{Nasr et al.~(2021)~\cite{nasr2021adversary}} & \myspan{DP-SGD} & \myspan{\approxdp} & \myspan{Poisson} & \myspan{MI} & \myspan{FPR/FNR} & \xmark & \cmark & \xmark \\
        & & & & & & \xmark & \xmark & \cmark \\
        \myrule
        Malek et al.~(2021)~\cite{malek2021antipodes} & \makecell{PATE-FM\\ALIBI} & Label DP & -- & AI & Accuracy & \cmark & \cmark & \xmark \\
        \myrule
        Lu et al.~(2022)~\cite{lu2022general} & \makecell{Naive Bayes\\Random Forest\\DP-SGD} & \approxdp & Poisson & MI & FPR/FNR & \xmark & \cmark & \xmark \\
        \myrule
        Houssiau et al.~(2022)~\cite{houssiau2022tapas} & \makecell{CTGAN\\MST\\PrivBayes} & \approxdp & -- & AI & FPR/FNR & \xmark & \cmark & \xmark \\
        \myrule
        Niu et al.~(2022)~\cite{niu2022dp} & \makecell{Report Noisy Max\\NoisyHist\\SVT\\Laplace\\RAPPOR} & \approxdp & -- & -- & Output-set & \xmark & \cmark & \cmark \\
        \myrule
        Tram{\`e}r et al.~(2022)~\cite{tramer2022debugging} & \makecell{Backpropagation Clipping} & \approxdp & Poisson & MI & FPR/FNR & \xmark & \cmark & \xmark \\
        \myrule
        Askin et al.~(2022)~\cite{askin2022statistical} & \makecell{Report Noisy Max\\SVT\\Laplace\\Exponential Mechanism} & \puredp & -- & -- & Distance Est. & \xmark & \cmark & \cmark \\
        \myrule
        Lokna et al.~(2023)~\cite{lokna2023group} & \makecell{MST\\Gaussian\\Discrete Gaussian\\Laplace} & \approxdp & -- & DPD~ & Output-set & \xmark & \cmark & \cmark \\
        \myrule
        Zanella-B{\'e}guelin et al.~(2023)~\cite{zanella2023bayesian} \hspace{-0.5cm} & \makecell{DP-SGD} & \approxdp & Poisson & MI & FPR/FNR & \xmark & \cmark & \xmark \\
        \myrule
        \myspan{Nasr et al.~(2023)~\cite{nasr2023tight}} & \myspan{DP-SGD} & \myspan{\fdp} & Poisson & \myspan{MI} & \myspan{FPR/FNR} & \xmark & \cmark & \xmark \\
        & & & -- & & & \cmark & \xmark & \cmark \\
        \myrule
        Gorla et al.~(2023)~\cite{gorla2023on} & \makecell{Truncated Laplace} & \rdp & -- & -- & Histogram Est. & \xmark & \cmark & \cmark \\
        \myrule
        Maddock et al.~(2023)~\cite{maddock2023canife} & DP-FedSGD & User-level DP & Poisson & MI & FPR/FNR & \cmark & \xmark & \xmark \\
        \myrule
        \myspan{Pillutla et al.~(2023)~\cite{pillutla2023unleashing}} & \myspan{DP-SGD} & \myspan{\approxdp} & \myspan{Poisson} & \myspan{MI} & \myspan{Custom} & \xmark & \cmark & \xmark \\
        & & & & & & \xmark & \xmark & \xmark \\
        \myrule
        \myspan{Steinke et al.~(2023)~\cite{steinke2023privacy}} & \myspan{DP-SGD} & \myspan{\approxdp} & \myspan{Poisson} & \myspan{MI} & \myspan{Accuracy} & \cmark & \cmark & \xmark \\
        & & & & & & \cmark & \xmark & \xmark \\
        \myrule
        Galen et al.~(2024)~\cite{galen2024oneshot} & DP-FedAvg & User-level DP & Shuffle & MI & Variance Est. & \cmark & \xmark & \xmark \\
        \myrule
        Matsumoto et al.~(2024)~\cite{matsumoto2024measuring} & \makecell{LDP-SGD} & \puredp & Shuffle & MI & FPR/FNR & \xmark & \xmark & \xmark \\
        \midrule
        \newpage
        \midrule
        & & \multicolumn{2}{c}{{\bf Foundations}} & \multicolumn{2}{c}{{\bf Operational Details}} & \multicolumn{3}{c}{{\bf Progress}} \\
        \cmidrule(lr){3-4}        \cmidrule(lr){5-6}         \cmidrule(lr){7-9}
        {\bf Reference} & \makecell{{\bf Mechanisms}\\{\bf Audited}} & {\bf DP Guarantee~} & {\bf Sub-Sampling} & {\bf Attack} & \makecell{{\bf Eval.}\\{\bf Function}} & {\bf Efficient~} & {\bf End-to-End~} & {\bf Tight} \\
        \midrule
        Chadha et al.~(2024)~\cite{chadha2024auditing} & \makecell{PATE\\CaPC\\PromptPATE\\Private k-NN} & \rdp & -- & MI & Custom & \xmark & \xmark & \xmark \\
        \myrule
        Zhang et al.~(2024)~\cite{zhang2024dp} & \makecell{Report Noisy Max\\NoisyHist\\SVT\\Laplace\\RAPPOR\\Prefix Sum\\Truncated Geometric} & \puredp & -- & DPD & Output-set & \xmark & \cmark & \cmark \\
        \myrule
        Chida et al.~(2024)~\cite{chida2024experimental} & \makecell{PrivBayes\\DPCopula\\Randomized Response} & MI & -- & MI & FPR/FNR & \xmark & \cmark & \xmark \\
        \myrule
        Lu et al.~(2024)~\cite{lu2024eureka} & \makecell{Report Noisy Max\\NoisyHist\\SVT} & \rdp & -- & DPD & Output-set & \xmark & \cmark & \cmark \\
        \myrule
        Xiang et al.~(2024)~\cite{xiang2024preserving} & \makecell{DP-GNN} & \fdp & Poisson & MI & FPR\FNR & \cmark & \xmark & \cmark \\
        \myrule
        \multirow{3}{*}{Annamalai et al.~(2024)~\cite{annamalai2024what}} & \multirow{3}{*}{\makecell{PrivBayes\\CTGAN\\DP-WGAN}} & \multirow{3}{*}{\approxdp} & \multirow{3}{*}{Poisson} & \multirow{3}{*}{MI} & \multirow{3}{*}{FPR/FNR} & \multirow{2}{*}{\xmark} & \multirow{2}{*}{\cmark} & \multirow{2}{*}{\xmark} \\
        & & & & & & & & \\
        & & & & & & \xmark & \xmark & \cmark \\
        \myrule
        Feng et al.~(2024)~\cite{feng2024privacy} & DP-SGD & \approxdp & Poisson & -- & Output-set & \xmark & \cmark & \cmark \\
        \myrule
        Annamalai et al.~(2024)~\cite{annamalai2024s} & \makecell{DP-SGD} & \fdp & Poisson & MI & FPR\FNR & \xmark & \xmark & \cmark \\
        \myrule
        Annamalai et al.~(2024)~\cite{annamalai2024nearly} & DP-SGD & \gdp & -- & MI & FPR/FNR & \xmark & \cmark & \xmark \\
        \myrule
        Yoon et al.~(2024)~\cite{yoon2024optimizing} & \makecell{DP-SGD} & \gdp & -- & MI & FPR/FNR & \xmark & \cmark & \xmark \\
        \myrule
        Kong et al.~(2024)~\cite{kong2024dp} & \makecell{Laplace\\Gaussian\\SVT\\DP-SGD} & \approxdp & -- & -- & Divergence Est. & \xmark & \xmark & \cmark \\
        \myrule
        Debenedetti et al.~(2024)~\cite{debenedetti2024privacy} & \makecell{DP-SGD} & \fdp & Poisson & MI & FPR/FNR & \xmark & \xmark & \xmark \\
        \myrule
        Arcolezi et al.~(2024)~\cite{arcolezi2024revealing} & \makecell{Generalized Randomized Response\\Subset Selection\\Local Hashing\\Unary Encoding\\Histogram Encoding} & \approxdp & -- & DPD & Output-set & \xmark & \cmark & \cmark \\
        \myrule
        \multirow{2}{*}{Mahloujifar et al.~(2024)~\cite{mahloujifar2024auditing}} & DP-SGD & \fdp & Poisson & \multirow{2}{*}{Recon~} & \multirow{2}{*}{Accuracy}  & \cmark & \xmark & \xmark \\
        & Gaussian & \gdp & -- &  &  & \cmark & \cmark & \cmark \\
        \myrule
        Annamalai et al.~(2024)~\cite{annamalai2024shuffle} & DP-SGD (Shuffle) & \approxdp & Shuffle & MI & FPR/FNR & \xmark & \xmark & \xmark \\
        \myrule
        Koskela et al.~(2025)~\cite{koskela2025auditing} & \edit{\makecell{Laplace\\Gaussian\\DP-SGD\\DP-SGD}} & \edit{HS Div.} & \edit{\makecell{--\\Poisson\\Poisson\\Poisson}} & \edit{\makecell{--\\--\\MI\\MI}} & \edit{Density Est.} & \edit{\makecell{\xmark\\\xmark\\\xmark\\\cmark}} & \edit{\makecell{\cmark\\\cmark\\\cmark\\\xmark}} & \edit{\cmark} \\
        \myrule
        Huang et al.~(2025)~\cite{huang2025auditing} & \makecell{Laplace\\Gaussian\\DP-SGD} & \approxdp & Poisson & -- & Variance Est. & \xmark & \xmark & \cmark \\
        \myrule
        Nasr et al.~(2025)~\cite{nasr2025the} & \makecell{DP-SGD} & \fdp & Poisson & MI & FPR/FNR & \xmark & \cmark & \xmark \\
        \myrule
        Cebere et al.~(2025)~\cite{cebere2025tighter} & DP-SGD & \gdp & Custom & MI & FPR/FNR & \xmark & \cmark & \xmark \\
        \myrule
        Ganev et al.~(2025)~\cite{ganev2025elusive} & \makecell{PATE-GAN} & \approxdp & -- & MI & FPR/FNR & \xmark & \cmark & \xmark \\
        \myrule
        Askin et al.~(2025)~\cite{askin2025general} & \makecell{Laplace\\Gaussian\\Sub-Sampling\\DP-SGD} & \fdp & Poisson & DPD & Output-set & \xmark & \cmark & \xmark \\
        \myrule
        Panda et al.~(2025)~\cite{panda2025privacy} & DP-SGD & \approxdp & Poisson & MI & Accuracy & \cmark & \cmark & \xmark \\
        \myrule
		{Xiang et al.~(2025)~\cite{xiang2025privacy}} & \makecell{DP-SGD\\Gaussian} & \makecell{\fdp\\\gdp} & \makecell{Poisson\\--} & {MI} & {Accuracy} & \makecell{\cmark\\\cmark} & \makecell{\xmark\\\cmark} & \makecell{\xmark\\\cmark} \\
        \bottomrule
        \caption{Full summary of key prior work on DP auditing.}
\label{tab:summary_full}
    \end{longtable}
\end{footnotesize}
%

\end{document}